\documentclass[10pt,a4paper]{article}
\usepackage[utf8]{inputenc}
\usepackage{amsmath}
\usepackage{amsfonts}
\usepackage{amssymb}
\usepackage{dsfont}
\usepackage{graphicx}
\usepackage{epstopdf}
\usepackage{subfigure}
\usepackage{color}
\usepackage[left=2.5cm,right=2.5cm,top=2.5cm,bottom=2.5cm]{geometry}
\usepackage{hyperref}

\usepackage{algorithm, algpseudocode}

\title{Deep Joint Learning valuation of Bermudan Swaptions}
\author{Francisco Gómez Casanova$^1$, Álvaro Leitao$^{2,3,}$\footnote{Corresponding author.}, Fernando de Lope Contreras$^1$,\\ and Carlos Vázquez$^{3,4}$}
\date{%
    $^1$BBVA, Spain\\%
    $^2$Universitat Oberta de Catalunya (UOC), Spain\\%
    $^3$CITIC Research Centre, University of A Coruña, Spain\\%
    $^4$Department of Mathematics, University of A Coruña, Spain\\[3ex]%
    \today
}

\begin{document}

\maketitle

\begin{abstract}

This paper addresses the problem of pricing involved financial derivatives by means of advanced of deep learning techniques. More precisely, we smartly combine several sophisticated neural network-based concepts like differential machine learning, Monte Carlo simulation-like training samples and joint learning to come up with an efficient numerical solution. The application of the latter development represents a novelty in the context of computational finance. We also propose a novel design of interdependent neural networks to price early-exercise products, in this case, Bermudan swaptions. The improvements in efficiency and accuracy provided by the here proposed approach is widely illustrated throughout a range of numerical experiments. Moreover, this novel methodology can be extended to the pricing of other financial derivatives.

\end{abstract}

\section{Introduction}
    
    As in many areas of research requiring the use of intensive scientific computing tools to solve mathematical models for real problems, there is an increasing use of deep learning and Artificial Neuronal Networks (ANN) techniques to overcome the difficulties associated to the use of more traditional numerical methods. This is also the case of the area of computational finance, where mathematical models associated to a large variety of problems related to the pricing and risk management of more or less complex financial products need to be efficiently solved. Among the variety of financial products, the class of financial derivatives correspond to those that depend on the evolution of other financial products or instruments (underlying assets): such as, equities, bonds, interest rates, foreign exchange rate, commodities, etc. We consider early-exercise derivatives, which are a type of financial contracts that allows the holder to exercise it at specific predetermined dates (exercise opportunities) prior to its expiration. This pricing problem has been widely handled by ``classical'' (non-based on ANNs) methods, although multiple (computational) challenges remain posed specially for the relatively complex financial derivatives. The early-exercise derivative valuation can be formulated as a free-boundary problem associated to Partial Differential Equations (PDEs), where not only the price of the derivative but also the exercise and non exercise regions as well as the optimal exercise region must be identified, thus requiring the use of appropriate numerical methods, like in \cite{Jaillet, Carlos_98, forsyth_amer, BerNogVaz_2006, FarVaz_2005, ASV-2020}, for example. Further, as for many other financial contracts, Monte Carlo methods can be employed in this context, typically relying on dynamic programming and backward induction to obtain the optimal exercise policy. The most successful contributions following this approach are based on the combination with appropriate regression techniques \cite{longstaff2001, jain2015, leitao2015, jain2019, ASV-MC1-2019, arregui2023}. Other methodologies that can be found in the literature employ trees \cite{echchafiq2023}, integration \cite{hagan2004, zhu2018} or Fourier inversion \cite{fang2009} techniques. However, all the enumerated methods present important drawbacks (precision, curse of dimensionality, general applicability, multiple evaluations, etc.) to be efficiently and robustly used at practical level, specially in terms of computational cost, and prevent their utilisation for providing accurate prices at the reasonable times required by the industry \cite{huge2020, ebner2022}.
    
    This paper addresses the problem of pricing involved financial derivatives by means of advanced forms of deep learning and ANN. Although the proposed methodology can be applied to other products, we will focus on the pricing of the Bermudan Swaption, which is a derivative whose underlying is the interest rate swap. In the corresponding section, we will motivate the additional difficulties in the pricing of this early-exercise interest rate derivative. In order to avoid some of the drawbacks described above when using traditional numerical techniques, the ANN solutions have recently emerged as interesting alternatives, whose main advantage is precisely that they decouple the expensive computations (carried on in training phase) from the actual use. An already significant number of solutions based on ANNs for financial problems have been proposed in the last few years \cite{kumar2023}. Thus, ANNs have been applied for pricing all types of financial derivatives, with \cite{becker2020, huge2020} and without \cite{liu2019} early-exercise features, recovering implied volatilities \cite{horvath2021, liu2021}, solving valuation PDEs \cite{salvador2021, villarino2023}, among others \cite{gnoatto2023}. Still, there is much room to explore within this novel field in general and, in particular, regarding its application in the early-exercise pricing problem.
    
    For this work, we employ a sophisticated form of ANN inspired in the developments proposed in \cite{huge2020}, where the so-called \emph{Differential Machine Learning} concept was introduced. The general idea behind it is to enhance the approximation power of an ANN by incorporating the information of the labels' differentials (when available or easily computed). Thus, henceforth, we denote the here employed ANN as \emph{Differential Artificial Neural Network} (DANN). Also in \cite{huge2020}, aiming to gain efficiency in the DANN training phase, the authors propose the use of the so-called \emph{sampled payoffs} as labels, instead of ground truth prices. In the common pricing context, this means to generate a single Monte Carlo path of the underlying model variable and consider the highly noisy price computed with it as the label to be employed in the training phase. Then, an entire training set (with thousands or millions of samples) can be generated at the cost of a classical Monte Carlo simulation-based pricing method. Here, we thus perform a Monte Carlo simulation of the considered model with a suitable time discretization (depending on the product at hand) and compute the corresponding prices/cash flows to be used as labels. Unlike the original approach, where the authors generate the all the sampled payoffs under the same distribution (i.e. with the same model parameters), in this work we take that idea a step further. Thus, with the goal of covering most of the market situations (represented by the model parameters), we simulate each of the Monte Carlo paths with a different set of parameters, such that, every single sample represents a very noisy price for that particular setting. Then, the DANN trained with these labels is able to learn the derivative prices for a wide set of market configurations, those defined in the ranges of the training set. This makes a great difference with respect to any other ``classical'' methodologies where, in order to obtain several prices, the corresponding algorithm needs to be repetitively executed (multiplying the computational cost). Moreover, the presented approach can be somehow generalised (as we do in this work), making that each sampled payoff is computed by averaging a bunch of few Monte Carlo realisations which share the same distribution.

    On top of the aforementioned approach, we introduce a novel strategy which intents to deepen in the idea of providing more available information to the ANN with the aim of improving its training performance and therefore producing more accurate estimations at similar computational cost. This new development consists of incorporating related financial products to be estimated by the ANN whose ground truth is ``easy'' to obtain. Then, besides the output/label of the ANN that represents the value of interest, additional outputs/labels are considered. Ideally, these aside financial products must have a strong connection with the original product (depending on the same model and market parameters). For example, it is well known that a derivative with early-exercise features relies on some kind of linear combination of the European counterparts. As we will see, this is precisely the type of relation that we exploit in this work. The smart combination of this idea with the differential machine learning and the (generalised) sampled payoffs constitutes the main contribution of this work, entailing several ways of improving the estimations provided by the ANN-based solution proposed here.

    The approach described in the previous paragraph shares common points to what is called \emph{joint learning} (also known as multi-task learning). This is a powerful machine learning paradigm that aims to improve the performance of multiple related tasks by simultaneously learning them in a shared framework \cite{crawshaw2020}. Unlike traditional single-task learning, joint learning leverages the inherent correlations and dependencies between tasks (quantities of interest) to enhance the overall network's generalization ability. By jointly optimizing the ANN model on multiple outputs, it can effectively transfer knowledge and representations learned from one output to benefit others, leading to more robust and efficient solutions \cite{crawshaw2020}. Joint learning finds application in various domains, such as natural language processing \cite{zhang2022}, computer vision \cite{muller2022}, and speech recognition \cite{li2021}, where tasks are interconnected and mutually beneficial learning can yield significant performance gains. The flexibility and versatility of joint learning make it an increasingly popular approach in the machine learning community, as it promotes better utilization of data and ultimately paves the way for more intelligent and adaptable learning systems \cite{ruder2017}.

    In the context of the previously described original aspects of this work, another contribution comes from the use of the described DANN model combined with joint learning to solve the problem of pricing early-exercise derivatives. As mentioned, Bermudan swaptions are considered, a very liquid and popular product in the financial markets, that represents a paradigmatic example of interest rate derivative with early-exercise features. Moreover, its valuation is rather challenging due to its particular properties, especially this early-exercise opportunity at a finite set of prescribed dates. For this purpose, we propose a novel design of interdependent DANNs which, once trained, encapsulate the optimal early-exercise policy. More precisely, each of these DANNs estimates the value of the derivative at the corresponding exercise time opportunity, which is employed for the DANN at the previous exercise date. This idea can be understood as a large scale generalization of the classical regression-based methods, although recalling that here we do not have a set of Monte Carlo simulations for the same underlying (with the same parameters) but a bunch of single paths instead, the parameterizations of which are different. To the best of our knowledge, this is the first time that this particular structure of interdependent DANNs is introduced to solve the Bermudan derivatives valuation. Further, one final DANN makes use of the ones codifying the early-exercise policy and performs the actual price estimation, following a traditional supervised training but, again, employing noisy labels (and their differentials).

    The rest of the paper is organized as follows. The next Section \ref{sec:formulation} simply reviews and establish the formulation framework in terms of the mathematical models, financial products and pricing approaches. The proper description and outcomes of the ANN-based solution proposed here are presented in Section \ref{sec:ANN}. The obtained results with the application to Bermudan swaptions are shown in Section \ref{sec:results}. Finally, Section \ref{sec:conclusions} concludes this work.

\section{Problem formulation}\label{sec:formulation}

    This section is devoted to set both the financial models for the interest rate derivatives we will consider, as well as the mathematical formulation that are followed throughout the paper. In view of the addressed financial derivatives, the underlying stochastic factor is the interest rate. In order to describe the time evolution of the over-night interest rate, we employ the  well-known \emph{Linear Gauss Markov} (LGM) model proposed in \cite{hagan2002, hagan2004}. A brief description is presented, together with their most important properties and the assumptions made. However, note that the deep learning approach here presented applies regardless the chosen model. Next, we briefly introduce the mathematical formulation of the valuation problem of both the European and the Bermudan swaptions under the LGM model, where the analytical formula and an alternative definition are presented for each case, respectively.

    \subsection{Linear Gauss Markov model}
        
        As previously mentioned, the here considered short-rate dynamics is the one given by LGM model \cite{hagan2002}. Due to its tractability and advantageous properties, this model is highly appreciated and used in the industry. In particular, a risk neutral measure associated to suitable \emph{numeraire} is directly available, and there is a well-established connection with the celebrated Hull-White model \cite{hullwhite1990}. Furthermore, LGM model is rather simple (in its one-factor version), with a single state variable, the stochastic process $x$, which evolves according to the equation:
        \begin{equation}\label{eq:lgm}
            dx_t = \alpha(t)dW_t, \quad x_0 = 0,
        \end{equation}
        where $W$ represents a standard Brownian motion under the risk neutral measure associated to the given numeraire and the variance of $x_t$ is $\zeta(t) := \int_0^t \alpha^2(\tau)d\tau$. Note that $x$ is a martingale under this risk neutral measure.
        
        Then, based on the state variable $x$, the \emph{numeraire} can be defined as follows
        \begin{equation*}
            N(t, x_t) = \frac{1}{D(t)}\exp\left(H(t)x_t + \frac{1}{2}H^2(t)\zeta(t)\right),
        \end{equation*}
        where $D(t)$ denotes the discount factor for time $t$ (observed in the market) and $H(t)$ is a curve with a similar interpretation as the mean reversion in the Hull-White model. Following existing literature, we choose
        \begin{equation*}
            H(t) = \frac{1 - \exp(-\kappa t)}{\kappa},
        \end{equation*}
        which entails that $\kappa$ corresponds exactly to the Hull-White mean reversion. The function $\alpha(t)$ is also related with the Hull-White mean reversion and volatility (see \cite{hagan2002}, for the details).
    
        Under the aforementioned premises, the following formula for the value at time $t$ of the zero coupon bond with maturity $T$ can be derived (see \cite{hagan2002}),
        \begin{equation*}
            Z(t, x_t; T) = \frac{D(T)}{D(t)}\exp\left(-(H(T) - H(t))x_t - \frac{1}{2}(H^2(T) - H^2(t))\zeta(t)\right).
        \end{equation*}

    \subsection{Valuation of European swaptions}\label{sec:european}

        A swaption is an option on a swap. The swap is a highly traded interest rate derivative consisting of interchanging a series of future payments between two parties at some predefined dates, $T_i, i=1, \dots, M$. One party pays a fixed amount given a fixed rate (fixed leg) while the other party pays a variable amount which depends on the market evolution of a stochastic floating rate (floating leg). In its simple form, this product is often known as \emph{interest rate swap} (IRS). The IRS value\footnote{For simplicity, here we assume that the nominal is a unit of currency.} can be entirely formulated in terms of zero coupon bonds, namely,
        \begin{equation*}
            V_S(t, x_t) = \phi\left(Z(t, x_t; T) - Z(t, x_t; T_M) - K\sum_{i = 1}^M \Delta T_i Z(t, x_t; T_i)\right)
        \end{equation*}
        where $K$ is the fixed interest rate, $T$ is the contract inception date, $\Delta T_i = T_i - T_{i-1}$, and $\phi$ determines whether the value is ``seen'' from the point of view of the party paying (swaption payer) the fixed amount ($\phi = 1$) or the party receiving (swaption receiver) it ($\phi = -1$).

        Thus, the European\footnote{When the holder of the option can exercise it only at one specific time in the future.} swaption payoff, provided that the exercise/maturity date is $T$, is given by
        \begin{equation*}
            V_E(T, x_T) = \max\left(V_S(T, x_T), 0\right) = \max\left(\phi\left(1 - Z(T, x_T, T_M) - K\sum_{i = 1}^M \Delta T_i Z(T, x_T, T_i)\right), 0\right)
        \end{equation*}
        whose value at time $t < T$ reads
        \begin{equation*}
            V_E(t, x_t) = N(t, x_t)\mathbb{E} \left[\frac{\max\left(\phi\left(1 - Z(T, x_T, T_M) - K\sum_{i = 1}^M \Delta T_i Z(T, x_T, T_i)\right), 0\right)}{N(T, x_T)} \Bigg| \mathcal{F}_t\right],
        \end{equation*}
        where $\mathcal{F}_t$ represents the filtration at $t$.

        Under the LGM model, an analytical solution for European swaptions can be straightforwardly obtained. For the sake of brevity, only the final formula is provided (further details can be found in \cite{hagan2002}, for example), which is given by
        \begin{equation}\label{eq:IRS_European_t}
        \begin{aligned}
            V_E(t, x_t)  &= \phi Z(t, x_t, T)\mathcal{N}\left(-\phi\frac{y_T^*}{\sqrt{\zeta(T) - \zeta(t)}}\right) \\
                    &- \phi Z(t, x_t, T_M)\mathcal{N}\left(-\phi\frac{y_T^* + (H(T_M) - H(T))(\zeta(T) - \zeta(t))}{\sqrt{\zeta(T) - \zeta(t)}}\right) \\
                    &- \phi K \sum_{i=1}^M \Delta T_i Z(t, x_t, T_i)\mathcal{N}\left(-\phi\frac{y_T^* + (H(T_i) - H(T))(\zeta(T) - \zeta(t))}{\sqrt{\zeta(T) - \zeta(t)}}\right),
        \end{aligned}
        \end{equation}
        where $\mathcal{N}$ denotes the cumulative distribution function (CDF) of the standard normal distribution, $y_t = x_t + H(t)\zeta(t)$ and $y_T^*$ is the unique solution that makes the payoff break-even. At today's time, i.e., at $t=0$ with $x_0=0$, the expression above can be further simplified, resulting in
        \begin{equation}\label{eq:IRS_European}
        \begin{aligned}
            V_E(0, 0)  &= \phi D(T)\mathcal{N}\left(-\phi\frac{y_T^*}{\sqrt{\zeta(T)}}\right) \\
                    &- \phi D(T_M)\mathcal{N}\left(-\phi\frac{y_T^* + (H(T_M) - H(T))\zeta(T)}{\sqrt{\zeta(T)}}\right) \\
                    &- \phi K \sum_{i=1}^M \Delta T_i D(T_i)\mathcal{N}\left(-\phi\frac{y_T^* + (H(T_i) - H(T))\zeta(T)}{\sqrt{\zeta(T)}}\right).
        \end{aligned}
        \end{equation}

    \subsection{Valuation of Bermudan swaptions}
    
         As described, under the LGM model, the European swaptions are readily priced. However, the valuation of their Bermudan counterpart (the goal of this project) is much more challenging (no closed-form solution is available, not even under the LGM), where the use of numerical approximations is mandatory. The Bermudan-style derivatives allow to exercise the contract (the swaption in this case) at several agreed future times. In order to try to simplify the computation of the value of Bermudan swaptions, we instead consider the valuation of a related, but easier product to price, namely, the \emph{Cancellable} IRS. There is a direct connection between both financial instruments, defined by the following relationships:
        \begin{equation}\label{eq:IRS_Bermudan}
        \begin{aligned}
            V_C^{p} &= V_S^{p} - V_B^{p}, \\
            V_C^{r} &= V_S^{r} - V_B^{r}, \\
        \end{aligned}
        \end{equation}
        where $V_S$, $V_C$, and $V_B$ denote the IRS, the Cancellable IRS and the Bermudan swaption prices, respectively. The superscripts indicate whether the (underlying) swap is priced from the point of view of the payer (p) or the receiver (r). 
        
        The Cancellable IRS allows to cancel (at some predefined times) the underlying IRS contract whenever the holder position at that particular time instant is no longer beneficial. Mathematically, assuming (without loss of generality) that the cancellation opportunities coincide with the IRS payment dates, the price of the Cancellable IRS can be rewritten as
        \begin{equation}\label{eq:IRS_Cancellable}
        \begin{aligned}
            \frac{V_C^{p}(t,x_t)}{N(t,x_t)} &= \sup_{\tau \in 
            \{T_i/T_i>t\}} \mathbb{E} \left[\max\left(\frac{V_S^p(\tau, x_{\tau})}{N(\tau,x_\tau)}, 0\right) \right], \\
            \frac{V_C^{r}(t,x_t)}{N(t,x_t)} &= \sup_{\tau \in 
            \{T_i/T_i>t\}} \mathbb{E} \left[\max\left(\frac{V_S^r(\tau, x_{\tau})}{N(\tau,x_\tau)}, 0\right) \right],
        \end{aligned}
        \end{equation}
        which enables the use of dynamic programming and backward induction to determine the optimal cancellation policy and, then, solve the problem. This is typically addressed, specially in the financial derivatives pricing context, by means of the combination of Monte Carlo simulation and regression techniques (see the introduction for references).

    \subsection{Further details about the financial product} \label{sec:details}
    
        In this work, without any loss of generality, some assumptions/simplifications have been adopted:
            \begin{itemize}
                \item The underlying IRS is spot start (starting today) with payment frequency 1 year (in both legs) associated to the tenor structure $\{T_1,T_2.\dots,T_M\}$, and expiry in 10 years.
                
                \item For the Bermudan swaption (respectively for the Cancellable IRS), the first call date is 1 year after the spot date of the underlying IRS, and the frequency of callability is also 1 year, i.e., the call dates of the Bermudan swaption coincide with the payment dates of the IRS.
                
                \item When the Bermudan swaption is exercised, the payment corresponding to the exercise date is assumed to be already paid, i.e, it is not included in the series of future payments.

                \item The notional is assumed to be the unit.
                    
            \end{itemize}

\section{Deep learning for pricing Bermudan swaptions under LGM model}\label{sec:ANN}

    We now move to the proper goal of this work, namely, the valuation of Bermudan swaptions by Deep Learning/ANN approaches, recalling that an alternative, but related, product, i.e., the cancellable IRS, is priced instead (see \eqref{eq:IRS_Bermudan} and \eqref{eq:IRS_Cancellable}). As previously indicated, the employed base network structure is the DANN, which is combined with a Monte Carlo-like generation of ``noisy'' labels (the so-called \emph{sampled payoffs}), to come up with an algorithm to compute the value of the Cancellable IRS. This plain approach will be coupled with the concept of joint learning (see Section \ref{sec:joint_learning}) which, as we will see, provides a remarkable enhancement.

    As mentioned, the sampled payoffs, generated as Monte Carlo paths, are employed to feed/train the DANN. Its generation is performed in terms of the parameter ranges described in Section \ref{sec:generation}, while its corresponding differentials (required for training the DANN) are computed by Automatic Adjoint Differentiation (AAD) \cite{savine2018}. Each path/sample is simulated by considering a different set of model and market parameters, which then become inputs of the DANN. This means that the here proposed DANN provides price estimations not only for a particular model configuration or market situation but also for a much wider combination of those. This represents a major advantage with respect to classical methods (including Monte Carlo and PDEs) or even plain applications of the DANN (see the seminal work in \cite{huge2020}), where the solutions are computed given a particular setting, i.e. a re-computation or a retraining are required every time the parameters of the addressed problem change. Moreover, the \emph{price sensitivities (differentials with respect to the model parameters) are obtained for free, implicitly provided by the advanced DANN structure design. However, this generalized simulation of training data based on sampled payoffs is also the reason why the joint learning becomes necessary: the space of inputs turns to be so big and the training data is highly ``noisy'' that the DANN model} requires extra information to reach the optimal solution with a manageable training set size at a reasonable computational cost. In addition, the smart selection of the parameter ranges to cover the realistic market dynamics (and avoid the less likely ones) also becomes a crucial aspect, which will be treated in Section \ref{sec:generation}.

    \subsection{DANNs design for the Cancellable IRS}

        As many other early-exercise derivative, the pricing methodology of a Cancellable IRS consists of two parts. First, the cancellation policy needs to be determined following a backward induction, i.e., starting from the last cancellation opportunity and iteratively computing the optimal cancellation strategy. Second, the cancellation policy is employed on a newly generated set of Monte Carlo simulations (to avoid biased estimations), thus obtaining the price of the Cancellable IRS.

        In order to approximate the cancellation policy, we propose a sequence of interconnected DANNs where a concrete DANN, associated with a cancellation opportunity, is employed (once trained) to compute the labels of the subsequent DANN. This structured construction implies that the labels for the DANN corresponding with the first cancellation opportunity depends on the estimations of all the ``previous'' DANNs. This specific design gets inspiration from the classical regression-based methodologies. We denote the sequence of DANNs as \emph{Backward DANNs}. Once the cancellation policy is encapsulated in the Backward DANNs, they can be employed over a newly generated set of Monte Carlo scenarios to compute the ``noisy'' price (sampled payoff) at scenario level. Those samples become the labels of an additional DANN, called here \emph{Forward DANN}. Particularly, the Backward DANNs are utilised to determine whether it is worth to cancel or not the IRS, incorporated in the methodology through an indicator function.
        
        In Algorithms \ref{alg:backwardMC} and \ref{alg:forwardMC}, the idea described above is mathematically represented. More precisely, for each $k$-th simulated path, we denote by $\Theta_k$ the model parameters, by $X_k$ the path values of LGM state variable, by $N_k$ the path values of \emph{numeraire}, and by $CF_k$ the future cash flows. The functions $\Tilde{V}^t$ and $\Bar{V}^0$ represent the Backward DANNs and the Forward DANNs, respectively. The line marked by ``Approximation by DANN'' represents the actual network training phase for which all the Monte Carlo paths are employed to feed the DANN\footnote{The implementation is fully vectorized at time level so the values for each path are treated together in single vectors.}.
        \begin{algorithm}
        \caption{Cancellation policy by the Backward DANNs.}\label{alg:backwardMC}
            \begin{algorithmic}
            \Require $\Theta_k$, $X_k(t)$, $N_k(t, X_k(t))$ and $F_k(s, t, \Theta_k, X_k) = \frac{CF_k(s, \Theta_k, X_k(s))}{N_k(t, X_k(t))}$
            \Ensure $V_k^+ = 0$ and $\mathcal{I}_k = 1$
            \For{$m = M-1 \dots 1$}
                \State $V_k(t_m, \Theta_k, X_k(t_m)) = N_k(t_m, X_k(t_m))\left(F_k(t_m, t_{m + 1}, \Theta_k, X_k) + \mathcal{I}_k\cdot V_k^+\right)$
                \State $\Tilde{V}^{t_m}(\Theta_k, X_k(t_m)) \approx V_k(t_m, \Theta_k, X_k(t_m))$ \Comment{Approximation by DANN.}
                \State $\mathcal{I}_k = \mathbf{1}_{\{\Tilde{V}^t_m(\Theta_k, X_k(t_m))>0\}}$ \Comment{Cancellation indicator (using the DANN approximation).}
                \State $V_k^+ = \frac{V_k(t_m, \Theta_k, X_k(t_m))}{N_k(t_m, X_k(t_m))}$ \Comment{Update $V^+_k$ for the next iteration.}
            \EndFor
        \end{algorithmic}
        \end{algorithm}
        
        \begin{algorithm}
        \caption{Pricing Cancellable IRS by the Forward DANN.}\label{alg:forwardMC}
            \begin{algorithmic}
            \Require $\hat{\Theta}_k$, $\hat{X}_k(t)$, $F_k(s, t,\hat{\Theta}_k, \hat{X}_k) = \frac{CF_k(s, \hat{\Theta}_k, \hat{X}_k(s))}{N_k(t, \hat{X}_k(t))}$ and $\Tilde{V}_k(t)$
            \Ensure $V_k^0(\hat{\Theta}_k) = 0$ and $\mathcal{I}_k = 1$
            \For{$m = 1 \dots M-1$}
                \State $\mathcal{I}_k = \mathcal{I}_k\cdot \mathbf{1}_{\{\Tilde{V}_k(t_m, \hat{\Theta}_k, \hat{X}_k(t_m))>0\}}$ \Comment{Cancellation indicator.}
                \State $V^0(\hat{\Theta}_k) = V^0(\hat{\Theta}_k) + \mathcal{I}_k\cdot F_k(t_m, t_{m+1}, \hat{\Theta}_k, \hat{X}_k)$ \Comment{Update $V^0$.}
            \EndFor
            \State $\Bar{V}^0(\hat{\Theta}_k) \approx V_k^0(\hat{\Theta}_k)$ \Comment{Approximation by DANN.}
        \end{algorithmic}
        \end{algorithm}

    \subsection{Training set generation}\label{sec:generation}

        The first step for any ANN-based solution is the generation of the training set to be used to adjust the free coefficients (weights) of the network. For this purpose, two aspects have a paramount importance: the domain of the input space and the sampling distribution within that domain. The first point is determined by the problem at hand, so the domain in a derivative valuation context is often selected based on the observed/expected market behaviour (past, present and future) or experience. The second aspect is trickier since, although a uniform distribution seems often appropriate, an smart sampling taking into account other factors like regions of more interest or with more error might provide a significant prediction improvement of the ANN model (given a training time budget). Also the  relation between the inputs (model and market parameters) is an important aspect to be exploited, which can be employed to avoid regions that represent unrealistic financial situations. In general, the generation of the training set must incorporate any information which allows to explore only the regions of interest within the space domain.
        
        Having the aforementioned aspects in mind, we propose to generate the training data required in this work employing the following strategies:
        \begin{itemize}
        
            \item \textbf{Mean reversion}. The Hull-White mean reversion parameter, $\kappa$, is uniformly distributed between given upper and lower bounds, that is
            \begin{equation*}
                \kappa = \mathcal{U}(l_{\kappa}, u_{\kappa}).
            \end{equation*}
        
            \item \textbf{LGM volatility}. The LGM volatility, $\alpha$, is assumed to be a piecewise constant function of the time. Further, a dependency on the $\kappa$ parameter is often observed in the market. Thus, we intend to incorporate that component in the samples' generation. For that, we start by generating samples of a forward ``implied'' volatility $\Sigma_j$ for each of the prescribed volatility intervals $(t_{j-1}, t_j], \, j=1,\dots J$. In order to produce volatilities which follow a certain structure and, again, avoid unrealistic volatility patterns, we firstly introduce a parametric model, namely, the recognised Rebonato's parameterization \cite{rebonato2009}, whose definition is
            \begin{equation*}
                h(t) = \left(a + bt\right)\exp\left(-ct\right) + d
            \end{equation*}
            Then, the values for the implied volatilities are chosen as
            \begin{equation*}
                \Sigma_j = h\left(\frac{t_{j-1} + t_j}{2}\right),
            \end{equation*}
            i.e., we evaluate the parametric function in the center of the interval of interest. To sample different volatility structures, the values of the Rebonato's model parameters can be randomly generated from uniform distributions within provided suitable ranges, i.e.,
            \begin{equation*}
                a = \mathcal{U}(l_a, u_a), \quad b = \mathcal{U}(l_b, u_b), \quad c = \mathcal{U}(l_c, u_c), \quad d = \mathcal{U}(l_d, u_d).
            \end{equation*}

            Once the implied volatilities are generated, the forward volatilities (the ones actually employed in the simulation) are calculated under a prescribed relation with the model parameter $\kappa$.
            Assuming we are in the Hull-White world, we take advantage of the equality
            \begin{equation*}
                \Sigma_j^2 \Delta t_j = \alpha_j^2 \int_{t_{j-1}}^{t_j} \exp\left(-2\kappa(t_j - s)\right) ds.
            \end{equation*}
            After solving the integral and some algebraic manipulations, we obtain that
            \begin{equation*}
                \alpha_j^2 = 2\kappa\frac{\Sigma_j^2 \Delta t_j}{1 - \exp(-\kappa\Delta t_j)},
            \end{equation*}
            where $\Delta t_j = t_j - t_{j-1}$ is the size of the $j-$th interval in the piecewise constant volatility function, which is given by
            \begin{equation*}
                \alpha(t) = \left\{
                \begin{array}{cc}
                    \alpha_1, & t \in (t_0, t_1], \\
                    \alpha_2, & t \in (t_1, t_2], \\
                    \alpha_3, & t \in (t_2, t_3], \\
                    \alpha_4, & t \in (t_3, t_4], \\
                    \vdots & \vdots \\
                    \alpha_J, & t \in (t_{J-1}, t_J].
                \end{array}
                \right.
            \end{equation*}

            In Figure \ref{fig:Alpha_hist}, an example of distributions of volatilities (generated with the prescribed values for the Test Case III in Section \ref{sec:base}) is presented, where only the first four values of the piecewise function (those that will be employed in this work) are shown. We observe that the samples are concentrated in the region of interest, determined by both the Rebonato's parameterization and the mean reversion, $\kappa$.
            \begin{figure}[h!]
                \centering
                \subfigure[$\alpha_1$.]{\includegraphics[width=0.42\textwidth]{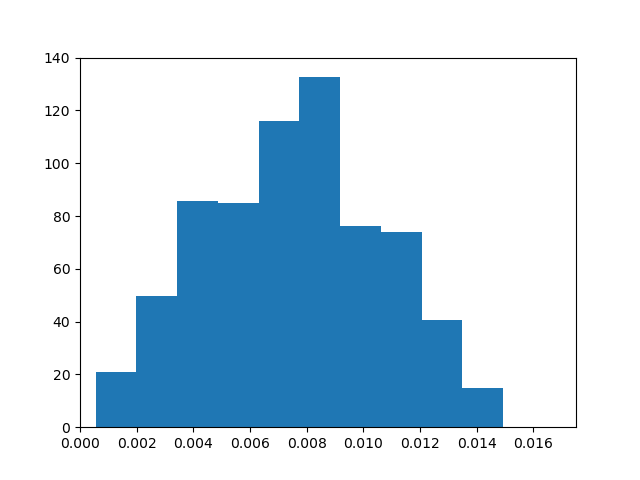}}
                \subfigure[$\alpha_2$.]{\includegraphics[width=0.42\textwidth]{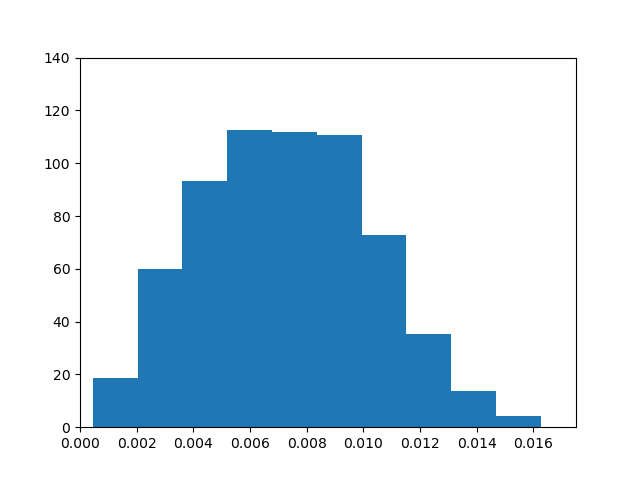}}
                \subfigure[$\alpha_3$.]{\includegraphics[width=0.42\textwidth]{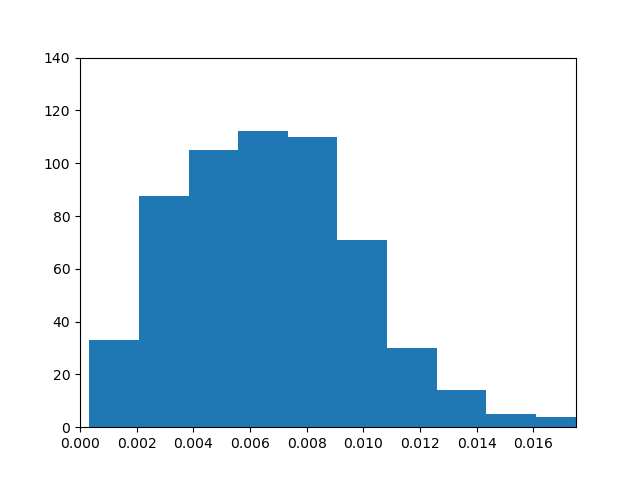}}
                \subfigure[$\alpha_4$.]{\includegraphics[width=0.42\textwidth]{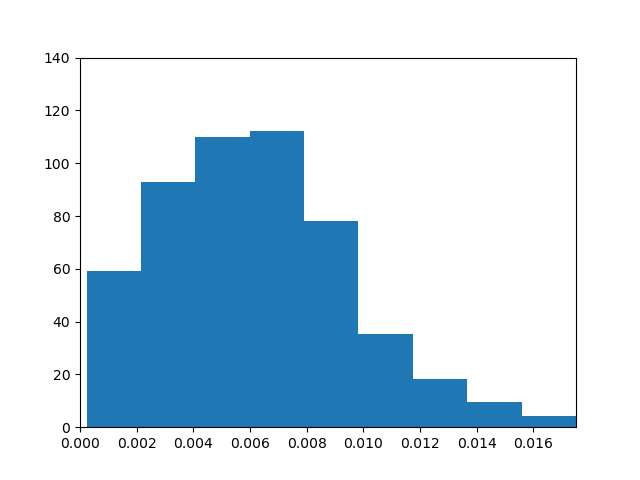}}
                \caption{Histograms of the volatilites.}
                \label{fig:Alpha_hist}
            \end{figure}

            \item \textbf{Discount factors}. Instead of directly addressing the discount curves, we will work with interest rate curves. For that, the well-known Nelson-Siegel parametric model \cite{nelson1987} is employed. Then, the model parameters are sampled to obtain the required number of interest rate curves and, subsequently, recover the corresponding discount factor curves. The model definition reads
            \begin{equation*}
                R(t) = \beta_0 G_0(t) + \beta_1 G_1(t) + \beta_2 G_2(t),
            \end{equation*}
            where
            \begin{equation*}
            \begin{aligned}
                G_0(t) &= 1, \\
                G_1(t) &= \frac{1 - \exp\left(-\frac{t}{\tau}\right)}{\frac{t}{\tau}}, \\
                G_2(t) &= \frac{1 - \exp\left(-\frac{t}{\tau}\right)}{\frac{t}{\tau}} - \exp\left(-\frac{t}{\tau}\right).
            \end{aligned}
            \end{equation*}
            The basis functions are not arbitrary, but represent the basic components of an interest rate term structure. The first basis function, $G_0$, represents the long-term level. The second function, $G_1$, represents an exponential decay and allows the term structure to slope upwards (with $\beta_1 < 0$) or downwards (with $\beta_1 > 0$). The third function, $G_2$, produces a butterfly effect, i.e., $\beta_2 > 0$ will produce a hump and $\beta_2 < 0$ will produce a trough. Finally, the parameter $\tau$ determines the location of the hump or the trough, as well as its steepness.

            Then, the Nelson-Siegel model parameters are randomly generated by
            \begin{equation*}
                \beta_0 = \mathcal{U}(l_0, u_0), \quad \beta_1 = \mathcal{U}(l_1, u_1), \quad \beta_2 = \mathcal{U}(l_2, u_2), \quad \tau = \mathcal{U}(l_{\tau}, u_{\tau}).
            \end{equation*}
            With the aim of generating realistic and up-to-date discount factors, the ranges of parameters are selected relying on the current market situation.
        
            Once we have generated as many interest rate curves as desired, the equivalent discount factors are easily recovered from the discount curve constructed as
            \begin{equation*}
                D(t) = \exp\left(-\int_0^t R(s)ds\right).
            \end{equation*}

            Next, in Figure \ref{fig:DF_hist}, we present an example of distributions of discount factors at different time points, namely, $t=1$, $t=4$, $t=7$, and $t=10$ (generated with the values prescribed for the Test Case III, see Section \ref{sec:base}).
            \begin{figure}[h!]
                \centering
                \subfigure[$t=1$.]{\includegraphics[width=0.42\textwidth]{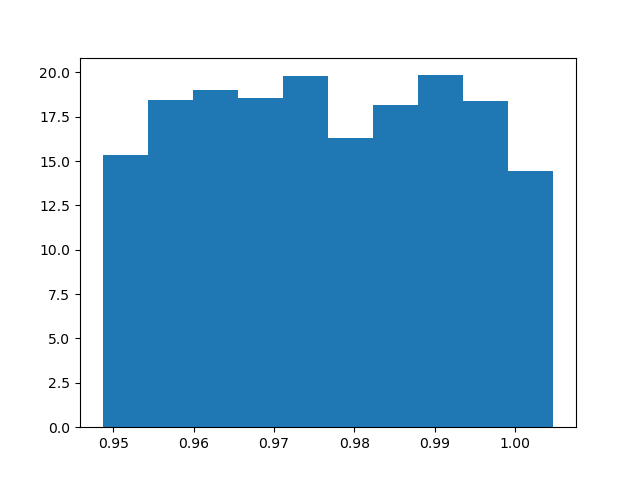}}
                \subfigure[$t=4$.]{\includegraphics[width=0.42\textwidth]{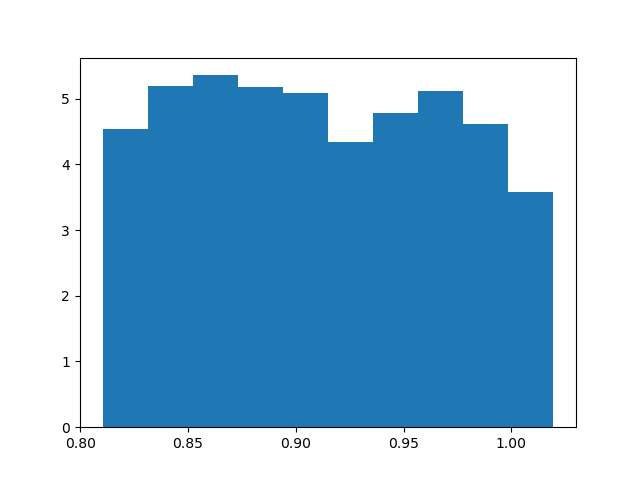}}
                \subfigure[$t=7$.]{\includegraphics[width=0.42\textwidth]{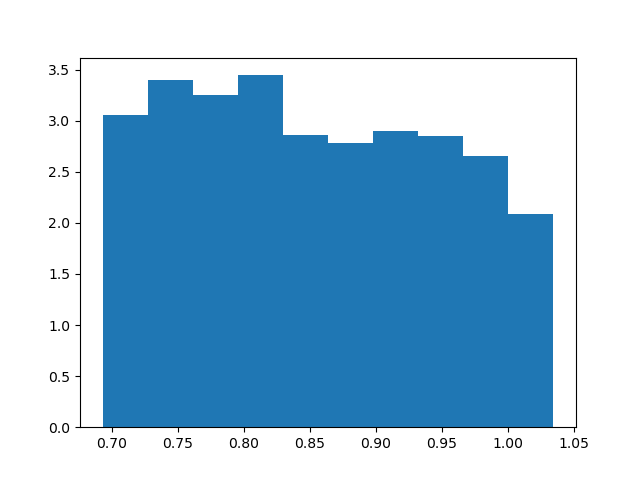}}
                \subfigure[$t=10$.]{\includegraphics[width=0.42\textwidth]{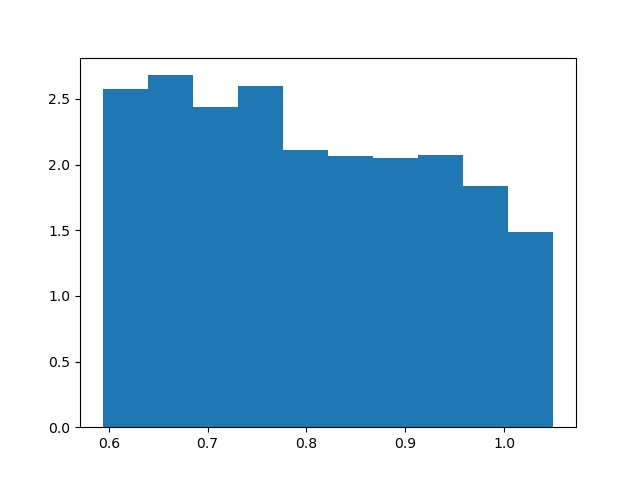}}
                \caption{Histograms of the discount factors at different time instants.}
                \label{fig:DF_hist}
            \end{figure}
    
            \item \textbf{Fixed rate}. The fixed rate of the underlying swap $K$, is uniformly perturbed from the ATM level, i.e., sampling the difference from the swap rate at time $t = 0$. Mathematically, we have
            \begin{equation*}
                K = ATM + \Delta K,
            \end{equation*}
            where
            \begin{equation*}
                ATM = \frac{1 - D(T_M)}{\sum_{i=1}^M\Delta T_i D(T_i)},
            \end{equation*} 
            with $\{T_i,\, i = 1,\dots, M\}$ being the set of swap payment times, and $\Delta K \in \mathcal{U}(l_{K}, u_{K})$ denotes the strike spread to be sampled.
        \end{itemize}

        The design of training data generation is straightforwardly translated to the input layer of the DANN where each entry represents one of the involved parameters in the training set definition. Thus, we then consider ten inputs representing the following parameters: $\kappa$, $a$, $b$, $c$, $d$, $\beta_0$, $\beta_1$, $\beta_2$, $\tau$, and $\Delta K$. For the sake of simplicity (and with certain abuse of notation), we indistinguishably employ the previous forms to refer both the parameters and the inputs.

    \subsection{Joint learning for Cancellable IRS: adding European swaptions as outputs}\label{sec:joint_learning}
        
        One of the main contributions of this work is the utilisation of the joint learning feature (in combination with the DANN) to significantly enhance the prediction power (accuracy) of the ANN-based solution. After the description of the proposed joint learning technique in this section, this enhancement will be illustrated in Section \ref{sec:joint}.

        In order to apply the joint learning, not only the Cancellable IRS is estimated by the ANN model, but also extra related outputs/labels are considered. In this case, a set of European swaptions whose maturities coincide with each of the cancellation times (what it is often called \emph{coterminal} swaptions) are chosen. As it is well-known that the Bermudan derivatives can be somehow seen as a combination of a number of their European counterparts, this represents a natural choice. Also note that the values of the European coterminal swaptions are often available in closed-form for many models in the literature. In particular, under the LGM model used here, their price is given by \eqref{eq:IRS_European}. This aspect is of great importance since, while the labels for the Cancellable IRS are noisy prices (sampled payoffs), the labels/prices of the coterminal European swaptions are the ground truth. Intuitively, adding labels with exact (non-noisy) values should help to improve the estimations provided by the whole DANN (at a prescribed training time budget), besides more quantities need to be predicted, boosted by the joint learning effect.

        Further, as has been mentioned throughout the paper, the joint learning approach is integrated within an already advanced ANN structure, namely, the DANN. This integration is not trivial as it entails the appearance of (cross) partial derivatives, implicating that the differential labels are no longer well-defined as 1D-vectors. Alternatively, they can be defined as directional differentials with respect to some specified linear combination of the outputs (see \cite{huge2020}, for details). By taking a special combination of directions in terms of unitary vectors, the Jacobian matrix is obtained, which facilitates the formal writing (in matrix form) of the backpropagation equations of the DANN\footnote{The multi-output backpropagation computation can be easily handled (and accelerated) by platforms like TensorFlow, which directly benefits from the CPU or GPU parallelism. Therefore, the additional computation complexity will be experienced as sublinear \cite{huge2020}.}. We then adapt our joint learning approach based on European coterminal swaptions by following the described differential strategy relying on the Jacobian matrices. This requires the computation of the coterminal swaptions differentials with respect to the involved parameters (i.e., network's inputs). Again, in this case, these differentials can be analytically computed, since their exact formula can be obtained just by differentiating in expression \eqref{eq:IRS_European}.

\begin{figure}[h!]
            \centering
            \includegraphics[width=0.95\textwidth]{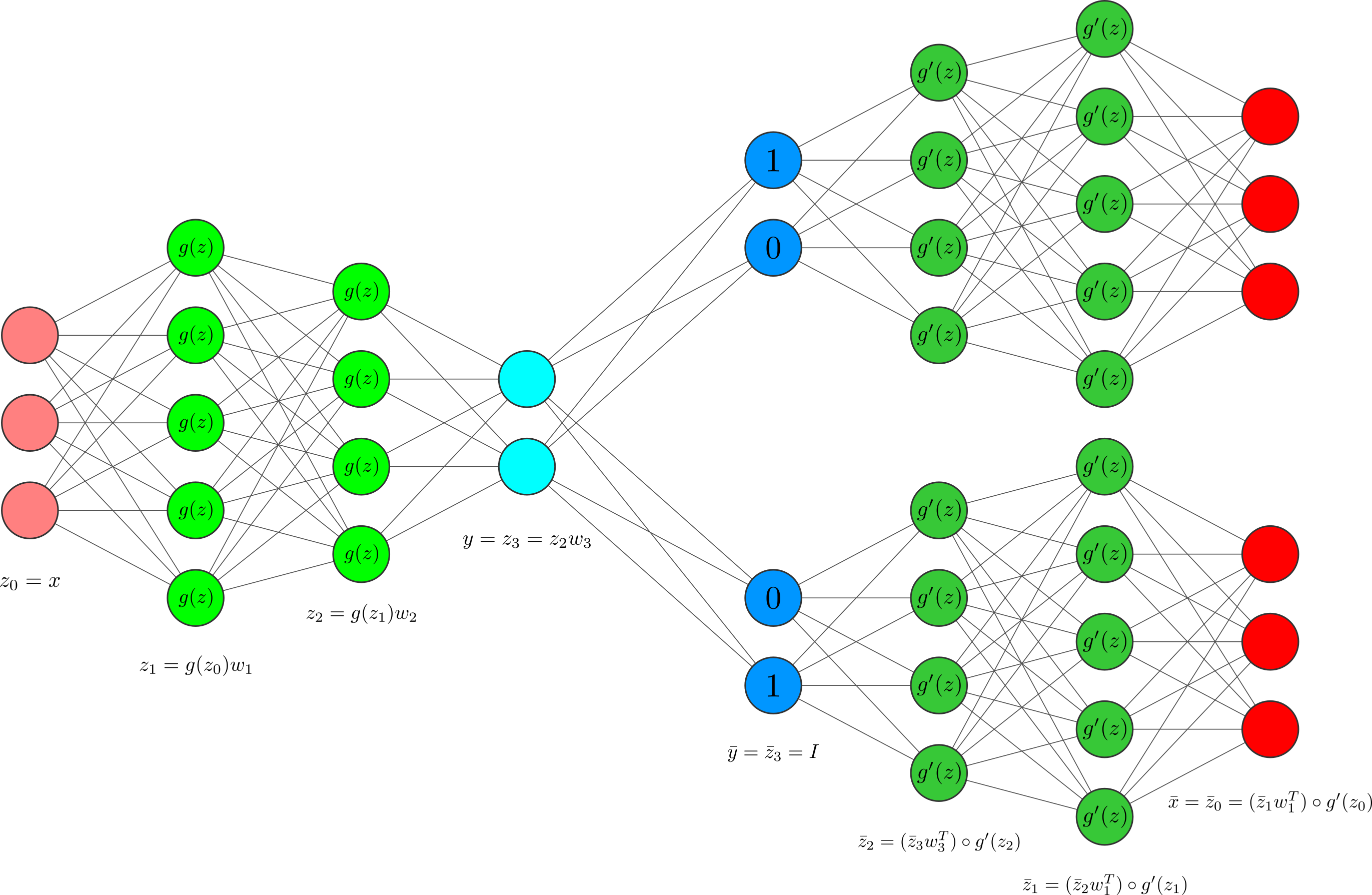}
            \caption{DANN structure considering multiple outputs, i.e., integrating the joint learning approach.}
            \label{fig:DANN}
        \end{figure}

        The final DANN structure, considering joint learning, is schematically represented in Figure \ref{fig:DANN}. This example considers a DANN with three inputs and two outputs. The plain feed-forward part is represented in lighter colors, while the differential (back-propagating) part is represented in darker ones. The values of the inputs and the outputs are denoted by the vectors $x$ and $y$, respectively. The intermediate results of the hidden layers are denoted by $z$ with a subscript identifying each layer. The feed-forward behaves as usual, ignoring the biases in the example for simplicity. Next, in the second part, the DANN model presents as many \emph{twin} networks as outputs in the feed-forward part (two in the example), introducing the differential component. This back-propagating part starts (following the base directional differentials) with the partial derivatives of the outputs with respect to themselves, thus resulting in the $0$'s and $1$'s which appear in the dark blue circles. Then, by applying the chain rule, the intermediate differentials are recursively calculated across the hidden layers (dark green circles) up to the  final ``differential'' layer (dark red circles), where the partial derivatives of the outputs with respect to the inputs are obtained. However, note that this procedure is computationally (and mathematically) treated as an integrated bigger network, handling the back-propagation via matrix operations which are made explicit at the bottom of each layer of neurons. The $\circ$ operator denotes the element-wise product and $I$ the identity matrix. As previously mentioned, by proceeding in this way, we obtain the Jacobian matrix as output of the whole differential part. Thus, the ``differential'' labels to be provided must be the partial derivatives with respect to each input, here denoted by means of the convenient \emph{adjoint} notation, namely, $\bar{x}$. Finally, we recall that the differential back-propagating network shares the weights with the feed-forward network. These weights are updated aiming to minimize a loss function that combines the outputs of both parts (feed-forward and differential) of the DANN model (see \cite{huge2020} for further details).

\section{Numerical experiments}\label{sec:results}

    The goal of this section is to measure the impact of different aspects in the accuracy of the predictions by the DANN. To do so, on the one hand, we follow an incremental procedure in terms of the involved inputs (see Section \ref{sec:base}) to have an intuition on the relevance of the different set of inputs. On the other hand, we aim to assess the gain provided by the proposed joint learning strategy (see Section \ref{sec:joint}). We complete the study by incorporating two additional features: the use of more Monte Carlo paths per sample (see Section \ref{sec:MCpaths}) and the use of extra temporal inputs to approximate the derivative price not only at the value date but also at any (prescribed) future time instant (see Section \ref{sec:time}).

    \subsection{Configuration}

        Regarding the hyperparameter configuration, after a systematic testing, we have chosen the structure and training design described in Table \ref{tab:hyperparameters}, which is the same for all the networks.
            \begin{table}[h!]
                \centering
                \begin{tabular}{c|c}
                    Hyperparameter & Value \\
                    \hline\hline
                    Layers & $4$ \\
                    Neurons & $32$ \\
                    Epochs & $128$ \\
                    Batch size & $4096$ \\
                    \hline
                \end{tabular}
                \caption{Hyperparameters.}
                \label{tab:hyperparameters}
            \end{table}

        The results included in this paper are mainly presented in the form of \emph{differences}' histograms. Those represent the distribution of distances between the reference values from a validation set and the values/prices predicted by the trained DANNs. In the captions of each histogram, the mean absolute error and the so-called \emph{interquantile interval} can be found. The latter is a sort of confidence interval, meaning that the $80\%$ of the predictions have a deviation that falls into the provided interval. The ground truth values (prices of Cancellable IRS) of the validation set are computed by a highly converged Monte Carlo pricer based on the so-called \emph{Stochastic Grid Bundling Method} (see \cite{jain2015}, for details) which ensures a high accuracy (below half basis point with $n=2^{20}$), as empirically verified\footnote{After a numerical study comparing the method against a Bermudan swaption pricer under the Hull-White model based on PDEs, implemented within the QuantLib library \cite{quantlib}.}. The size of the validation set employed here is $4096$.

        Other considerations:
        \begin{itemize}
            \item The number of Monte Carlo simulation time steps are taken equal to the swap tenor.

            \item The data of the piecewise constant volatility are $J=3$ and $t_0 =0, \, t_1=1, \, t_2=5$ and $t_3=10$.
            
            \item Random variables samples are generated by antithetic variables techniques to reduce variance.
            
            \item Differentials are obtained via the AAD module of the TensorFlow package.
            
            \item The codes have been implemented in Python 3.8 with TensorFlow 2.7 on the OS Linux Ubuntu 20.04, processor CPU Intel Core i7-4720HQ 2.6GHz, RAM memory of 16GB and GPU Nvidia Tesla V100. The computations are performed in single precision.
        \end{itemize}

        \subsubsection{Test base cases}\label{sec:base}
    
            As mentioned, in order to keep more control over the obtained approximations by the DANN, we establish incremental base test cases, which gives some substantial insights of how the estimations behave. To define the test cases, the inputs are grouped depending on what they represent: mean reversion ($\kappa$), volatility ($a$, $b$, $c$, and $d$), discount factors ($\beta_0$, $\beta_1$, $\beta_2$, and $\tau$), and strike spread ($\Delta K$). Then, each group of inputs is incrementally incorporated to the DANN training. Note that the input layer size is kept invariant ($10$ inputs), but a specific group of inputs can be disabled by reducing its sampling domain to the minimum such that the DANN treats that particular input as a constant. Thus, following this idea, the incremental test cases are constructed (recalling that, with those input parameters, a set of Monte Carlo paths is generated) as shown in Table \ref{tab:test_cases}.
            \begin{table}[h!]
            \centering
                \begin{tabular}{l|l}
                    \hline
                    Test Case I & Test Case II \\
                    \hline
                    $l_\kappa = -0.05, \quad u_\kappa = 0.1$ & $l_\kappa = -0.05, \quad u_\kappa = 0.1$ \\
                    $l_a = -10^{-5}, \quad u_a = 10^{-5}$ & $l_a = 10^{-5}, \quad u_a = 0.0075$ \\
                    $l_b = -10^{-5}, \quad u_b = 10^{-5}$ & $l_b = 0, \quad u_b = 0.0005$ \\
                    $l_c = -10^{-5}, \quad u_c = 10^{-5}$ & $l_c = 0, \quad u_c = 0.25$ \\
                    $l_d = 0.0075 - 10^{-5}, \quad u_d = 0.0075 + 10^{-5}$ & $l_d = 10^{-5}, \quad u_d = 0.0075$ \\
                    $l_0 = 0.02 - 10^{-5}, \quad u_0 = 0.02 + 10^{-5}$ & $l_0 = 0.02 - 10^{-5}, \quad u_0 = 0.02 + 10^{-5}$ \\
                    $l_1 = -10^{-5}, \quad u_1 = 10^{-5}$ & $l_1 = -10^{-5}, \quad u_1 = 10^{-5}$ \\
                    $l_2 = -10^{-5}, \quad u_2 = 10^{-5}$ & $l_2 = -10^{-5}, \quad u_2 = 10^{-5}$ \\
                    $l_\tau = 1 - 10^{-5}, \quad u_\tau = 1 + 10^{-5}$ & $l_\tau = 1 - 10^{-5}, \quad u_\tau = 1 + 10^{-5}$ \\
                    $l_K = -10^{-5}, \quad u_K = 10^{-5}$ & $l_K = -10^{-5}, \quad u_K = 10^{-5}$ \\
                    \hline\hline
                    Test Case III & Test Case IV \\
                    \hline
                    $l_\kappa = -0.05, \quad u_\kappa = 0.1$ & $l_\kappa = -0.05, \quad u_\kappa = 0.1$ \\
                    $l_a = 10^{-5}, \quad u_a = 0.0075$ & $l_a = 10^{-5}, \quad u_a = 0.0075$ \\
                    $l_b = 0, \quad u_b = 0.0005$ & $l_b = 0, \quad u_b = 0.0005$ \\
                    $l_c = 0, \quad u_c = 0.25$ & $l_c = 0, \quad u_c = 0.25$ \\
                    $l_d = 10^{-5}, \quad u_d = 0.0075$ & $l_d = 10^{-5}, \quad u_d = 0.0075$ \\
                    $l_0 = -0.005, \quad u_0 = 0.05$ & $l_0 = -0.005, \quad u_0 = 0.05$ \\
                    $l_1 = 0, \quad u_1 = 0.001$ & $l_1 = 0, \quad u_1 = 0.001$ \\
                    $l_2 = 0, \quad u_2 = 0.01$ & $l_2 = 0, \quad u_2 = 0.01$ \\
                    $l_\tau = 0.01, \quad u_\tau = 2$ & $l_\tau = 0.01, \quad u_\tau = 2$ \\
                    $l_K = -10^{-5}, \quad u_K = 10^{-5}$ & $l_K = -0.01, \quad u_K = 0.01$ \\
                    \hline
                \end{tabular}
                \caption{Test cases.}
                \label{tab:test_cases}
            \end{table}
            The limits of the ranges for each parameter/input are selected such that a selected real market situation is well represented.

	\subsection{Impact of joint learning}\label{sec:joint}
	
\begin{figure}[h!]
			\centering
			\subfigure[Abs. avg.: $4.3$, $(Q_{10}, Q_{90}) = (-5.7, 7.0)$.]{\includegraphics[width=0.49\textwidth]{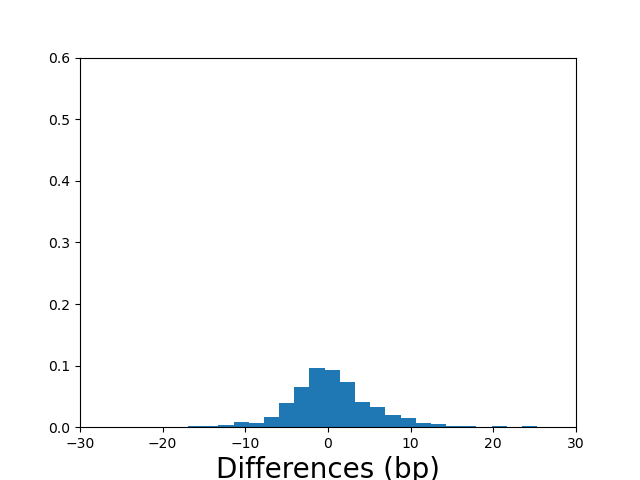}}
			\subfigure[Abs. avg.: $0.7$, $(Q_{10}, Q_{90}) = (-0.8, 1.4)$.]{\includegraphics[width=0.49\textwidth]{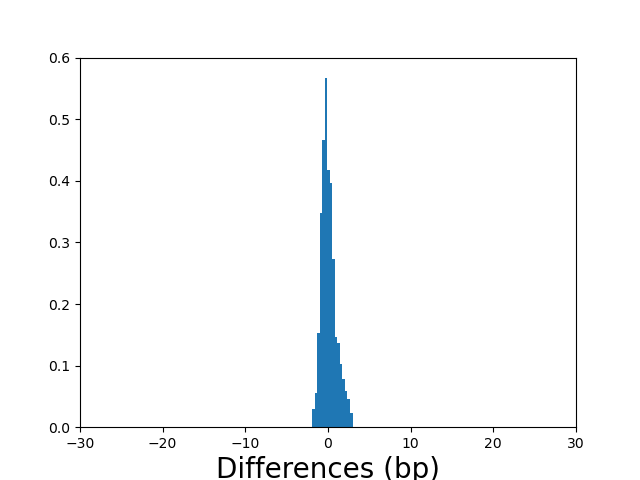}}
			\caption{Pricing differences in basis points of Test Case I: Plain DANN (left) and DANN with joint learning (right).}
			\label{fig:diferences_test_I}
		\end{figure}

\begin{figure}[h!]
			\centering
			\subfigure[Abs. avg.: $3.9$, $(Q_{10}, Q_{90}) = (-6.0, 5.7)$.]{\includegraphics[width=0.49\textwidth]{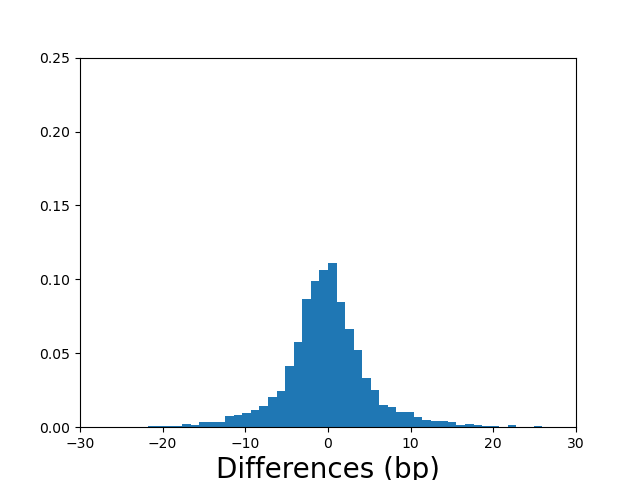}}
			\subfigure[Abs. avg.: $1.8$, $(Q_{10}, Q_{90}) = (-1.0, 3.7)$.]{\includegraphics[width=0.49\textwidth]{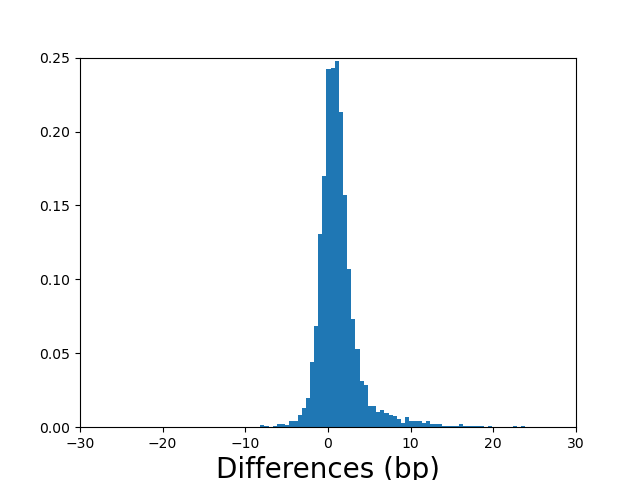}}
			\caption{Pricing differences in basis points of Test Case II: Plain DANN (left) and DANN with joint learning (right).}
			\label{fig:diferences_test_II}
		\end{figure}

\begin{figure}[h!]
			\centering
			\subfigure[Abs. avg.: $8.3$, $(Q_{10}, Q_{90}) = (-10.0, 12.4)$.]{\includegraphics[width=0.49\textwidth]{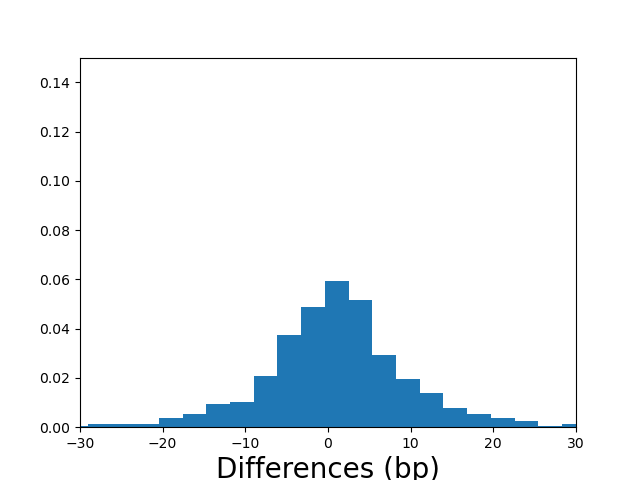}}
			\subfigure[Abs. avg.: $2.6$, $(Q_{10}, Q_{90}) = (-1.9, 5.1)$.]{\includegraphics[width=0.49\textwidth]{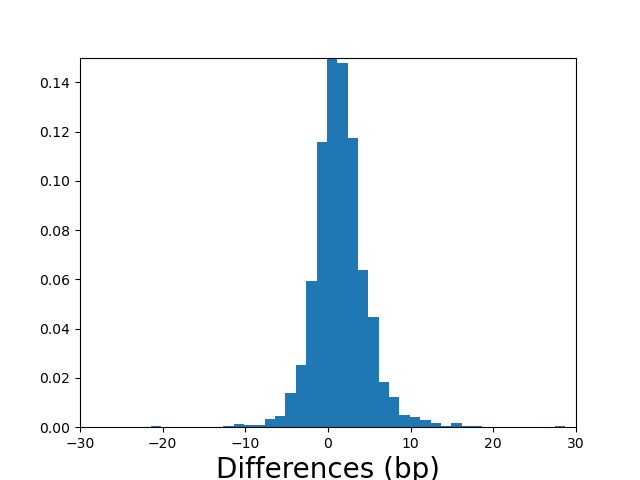}}
			\caption{Pricing differences in basis points of Test Case III: Plain DANN (left) and DANN with joint learning (right).}
			\label{fig:diferences_test_III}
		\end{figure}

\begin{figure}[h!]
			\centering
			\subfigure[Abs. avg.: $5.0$, $(Q_{10}, Q_{90}) = (-8.1, 7.4)$.]{\includegraphics[width=0.49\textwidth]{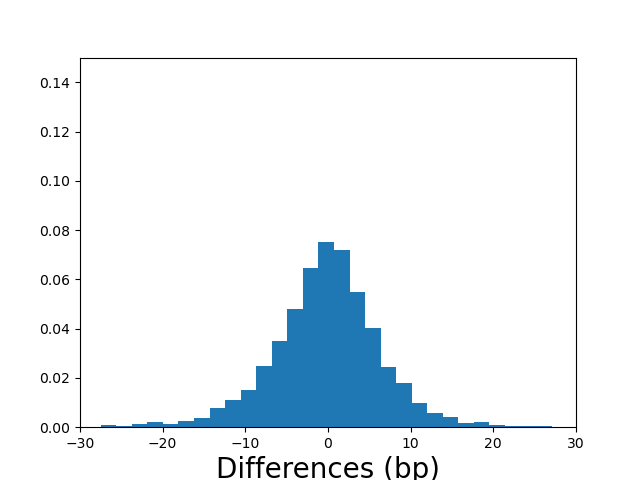}}
			\subfigure[Abs. avg.: $2.3$, $(Q_{10}, Q_{90}) = (-4.8, 2.5)$.]{\includegraphics[width=0.49\textwidth]{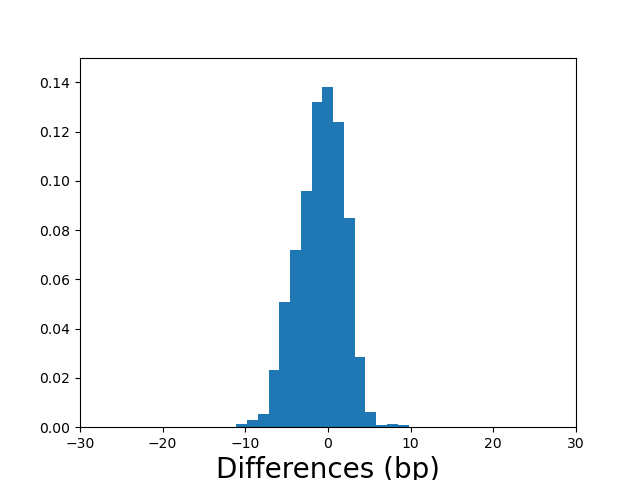}}
			\caption{Pricing differences in basis points of Test Case IV: Plain DANN (left) and DANN with joint learning (right).}
			\label{fig:diferences_test_IV}
		\end{figure}

		For each of the test cases defined above, we measure the actual performance of the whole ``system'' of DANNs in estimating the price of the Cancellable IRS. In the experiments of this section, the training set size is $n_b = 2^{22}$ for the Backward DANNs and also $n_f = 2^{22}$ for the Forward DANN. In the left graph of Figures \ref{fig:diferences_test_I}, \ref{fig:diferences_test_II}, \ref{fig:diferences_test_III} and, \ref{fig:diferences_test_IV}, the differences distributions are depicted. Next, we incorporate joint learning construction which, as mentioned, relies on the coterminal European swaptions. The results of the DANN estimations with joint learning are presented in the right graph of the same Figures \ref{fig:diferences_test_I}, \ref{fig:diferences_test_II}, \ref{fig:diferences_test_III} and, \ref{fig:diferences_test_IV}.

		From the results observed in the figures, we extract the following insights:
		\begin{itemize}
			\item In all cases the differences' distributions are centered at cero, thus indicating that the DANN predictions do not present bias.
			
			\item The incorporation of the inputs related with the discount factors and the volatility seems to make the solution to be approximated more challenging.

			\item When the strike spread is included (test case IV) the DANN provides slightly better estimations. Although this might seem counterintuitive, that behavior appears due to the effect of the strikes far from ATM level (particularly those more in-the-money). As we measure the absolute differences, the estimations of the DANN present a smaller deviation when the derivative price is lower. This effect is implicitly present in the histograms, where a certain skewness is observed (more clearly visible in the case employing joint learning, see Figure \ref{fig:diferences_test_IV}b).

			\item An impressive (and general) reduction of the error thanks to the joint learning approach is achieved, where both the average error and the interquantile interval are, at least, halved.
		\end{itemize}

	\subsection{Impact of number of samples and Monte Carlo paths per sample}\label{sec:MCpaths}

\begin{figure}[h!]
			\centering
			\subfigure[$(Q_{10}, Q_{90}) = (-8.9, 8.8)$.]{\includegraphics[width=0.32\textwidth]{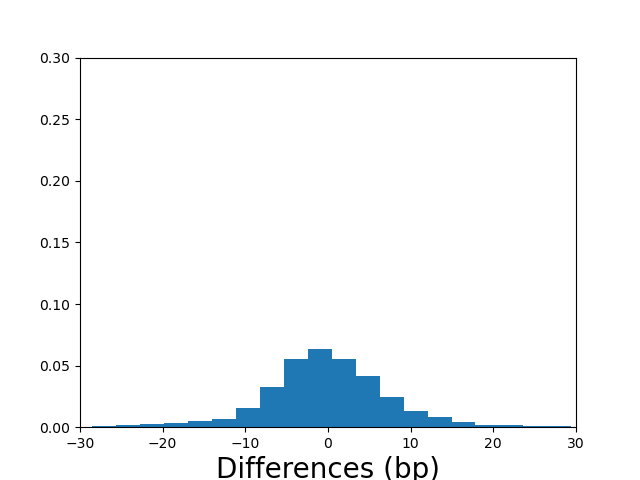}}
			\subfigure[$(Q_{10}, Q_{90}) = (-6.9, 5.9)$.]{\includegraphics[width=0.32\textwidth]{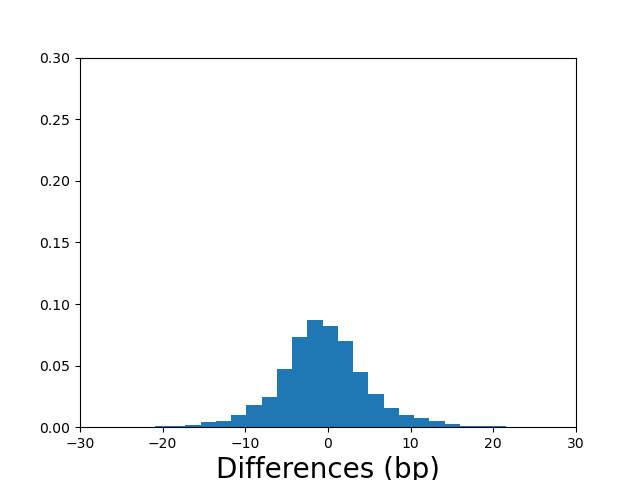}}
			\subfigure[$(Q_{10}, Q_{90}) = (-6.5, 3.4)$.]{\includegraphics[width=0.32\textwidth]{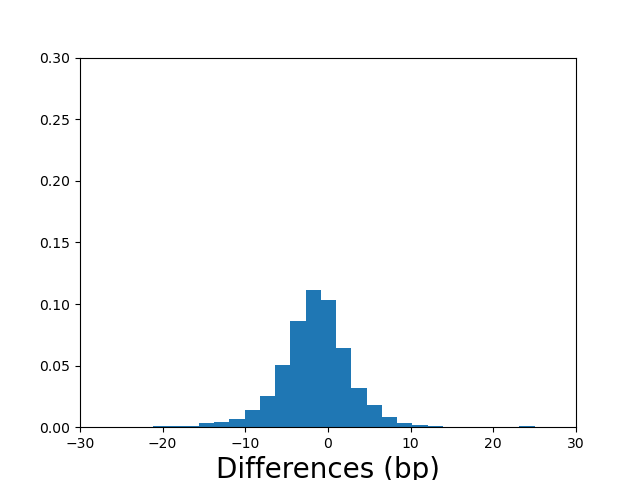}}
			\subfigure[$(Q_{10}, Q_{90}) = (-3.6, 5.2)$.]{\includegraphics[width=0.32\textwidth]{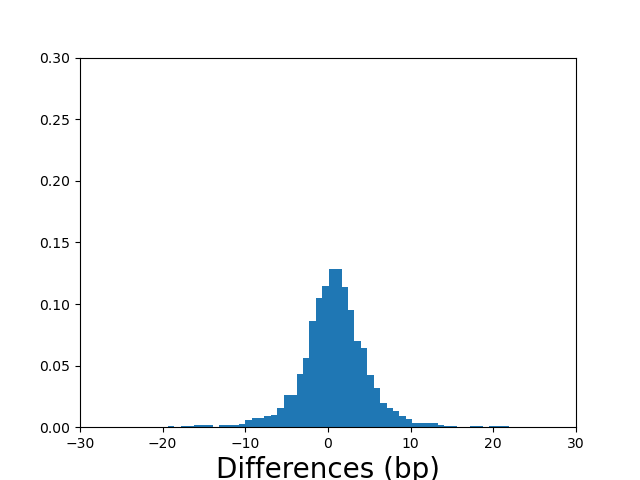}}
			\subfigure[$(Q_{10}, Q_{90}) = (-2.5, 4.8)$.]{\includegraphics[width=0.32\textwidth]{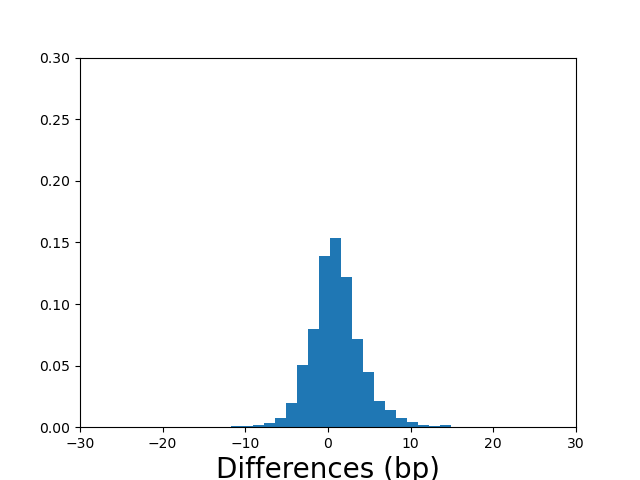}}
			\subfigure[$(Q_{10}, Q_{90}) = (-3.0, 3.2)$.]{\includegraphics[width=0.32\textwidth]{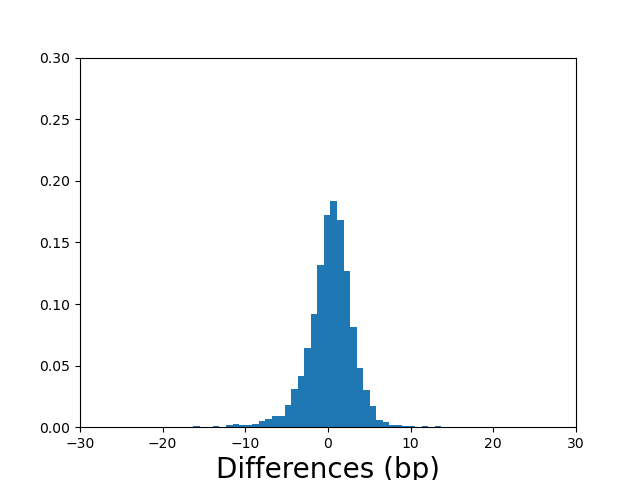}}
			\subfigure[$(Q_{10}, Q_{90}) = (-2.5, 2.7)$.]{\includegraphics[width=0.32\textwidth]{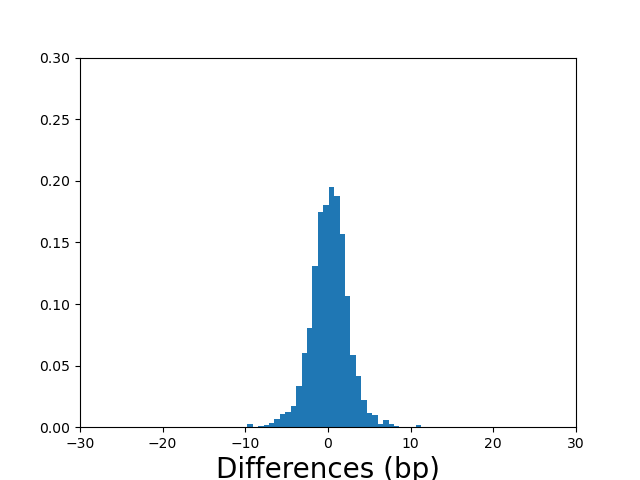}}
			\subfigure[$(Q_{10}, Q_{90}) = (-1.9, 2.7)$.]{\includegraphics[width=0.32\textwidth]{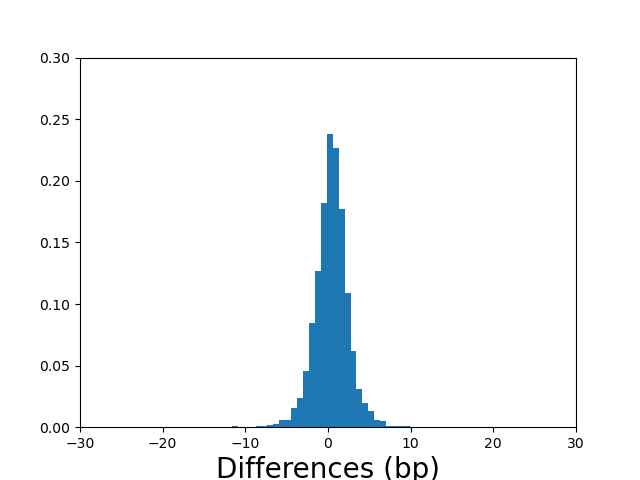}}
			\subfigure[$(Q_{10}, Q_{90}) = (-1.6, 2.3)$.]{\includegraphics[width=0.32\textwidth]{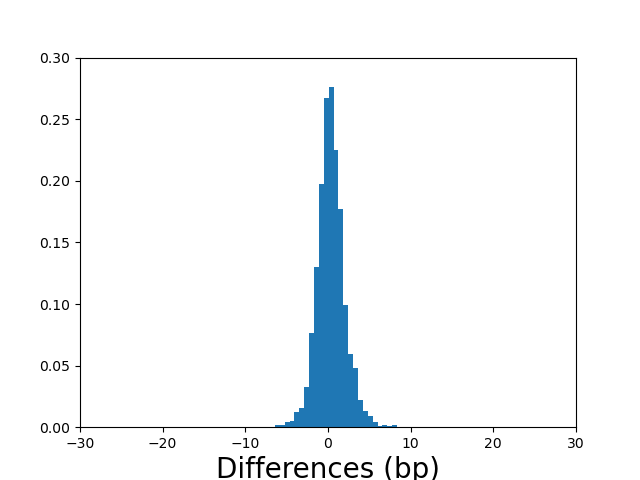}}
			\caption{Pricing differences in basis points. Test Case IV. Plain DANN (no joint learning). Columns: $n_f = 2^{22}$ (left), $n_f = 2^{23}$ (central), $n_f = 2^{24}$ (right); Rows: $n_{MC} = 1$ (top), $n_{MC} = 4$ (middle), $n_{MC} = 16$ (bottom).}
			\label{fig:impact_samples_mc}
		\end{figure}

\begin{figure}[h!]
			\centering
			\subfigure[$(Q_{10}, Q_{90}) = (-4.8, 2.7)$.]{\includegraphics[width=0.32\textwidth]{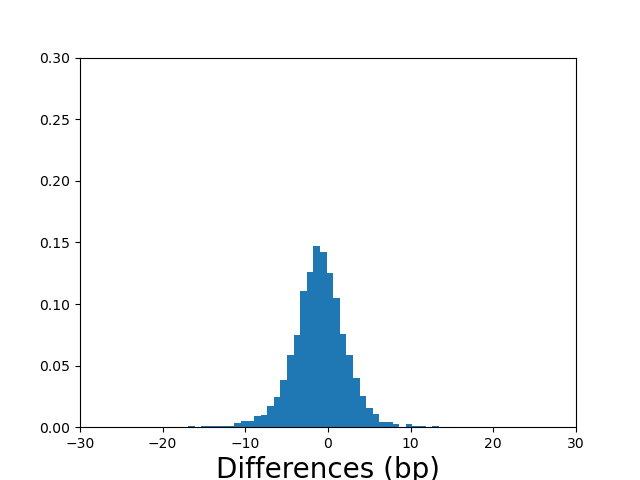}}
			\subfigure[$(Q_{10}, Q_{90}) = (-3.1, 2.9)$.]{\includegraphics[width=0.32\textwidth]{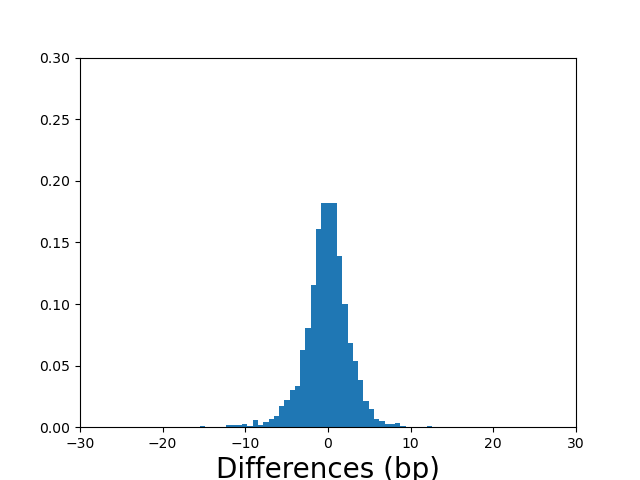}}
			\subfigure[$(Q_{10}, Q_{90}) = (-2.8, 2.8)$.]{\includegraphics[width=0.32\textwidth]{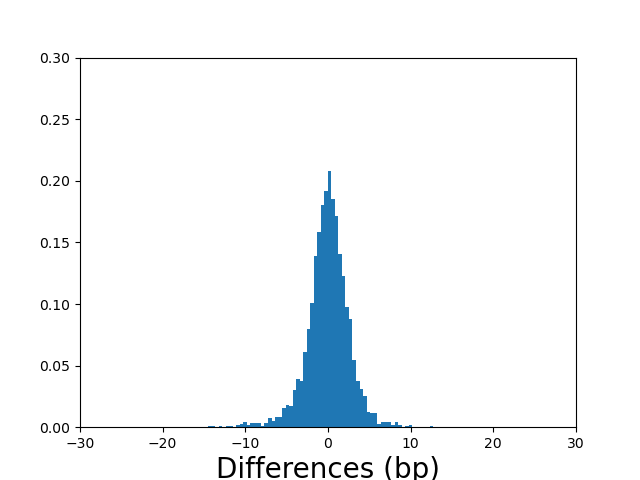}}
			\subfigure[$(Q_{10}, Q_{90}) = (-4.1, 1.3)$.]{\includegraphics[width=0.32\textwidth]{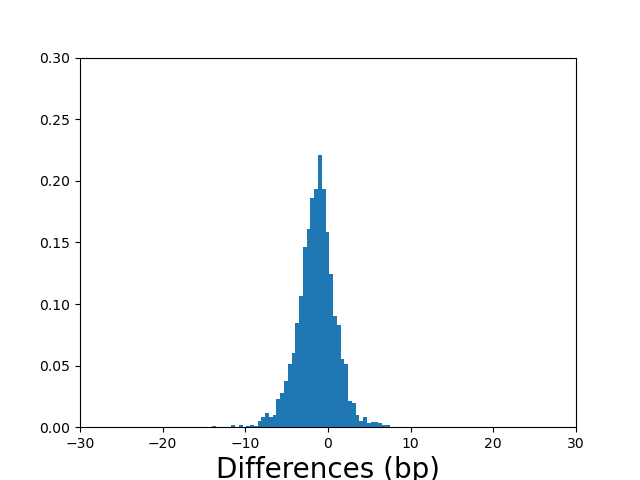}}
			\subfigure[$(Q_{10}, Q_{90}) = (-2.1, 3.0)$.]{\includegraphics[width=0.32\textwidth]{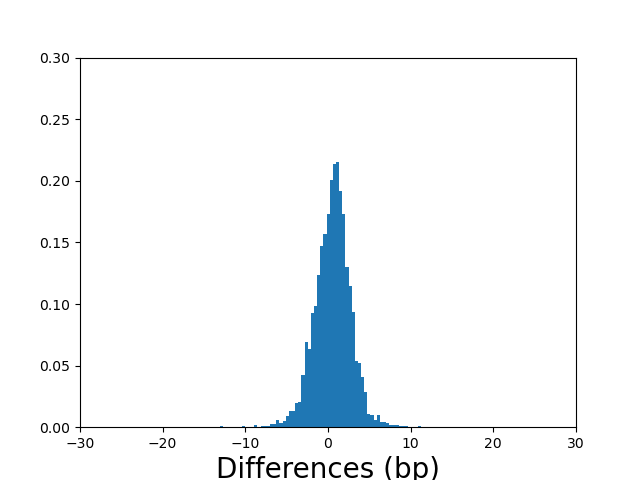}}
			\subfigure[$(Q_{10}, Q_{90}) = (-0.4, 2.9)$.]{\includegraphics[width=0.32\textwidth]{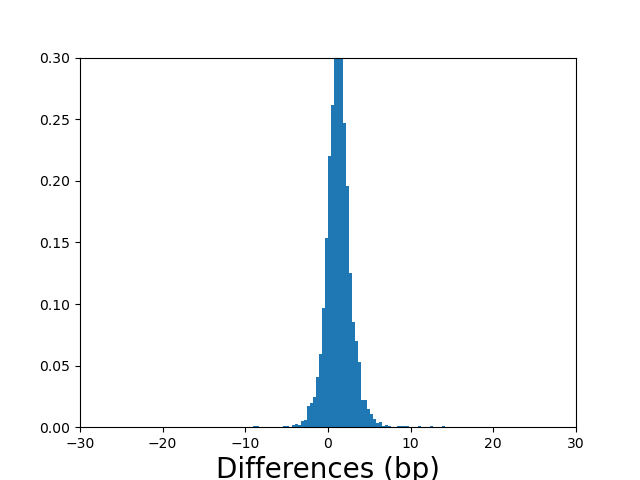}}
			\subfigure[$(Q_{10}, Q_{90}) = (-1.2, 3.0)$.]{\includegraphics[width=0.32\textwidth]{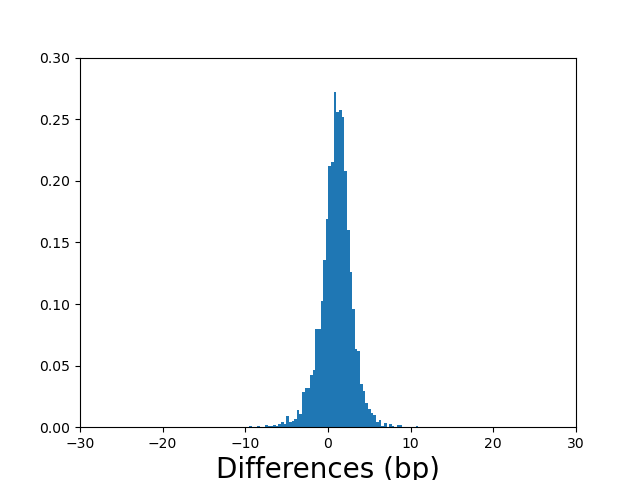}}
			\subfigure[$(Q_{10}, Q_{90}) = (-1.7, 2.1)$.]{\includegraphics[width=0.32\textwidth]{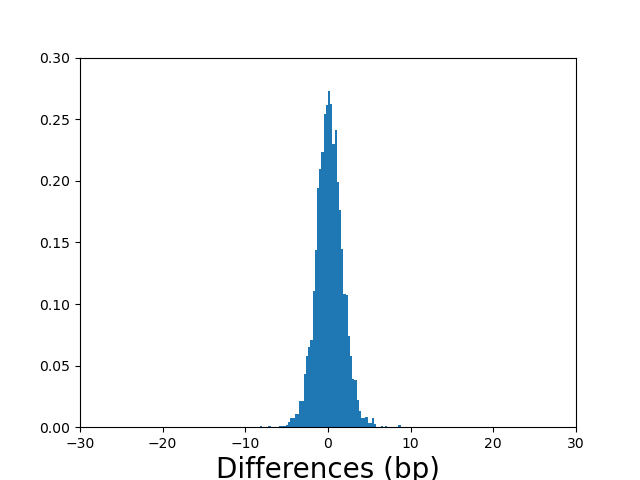}}
			\subfigure[$(Q_{10}, Q_{90}) = (-1.7, 2.1)$.]{\includegraphics[width=0.32\textwidth]{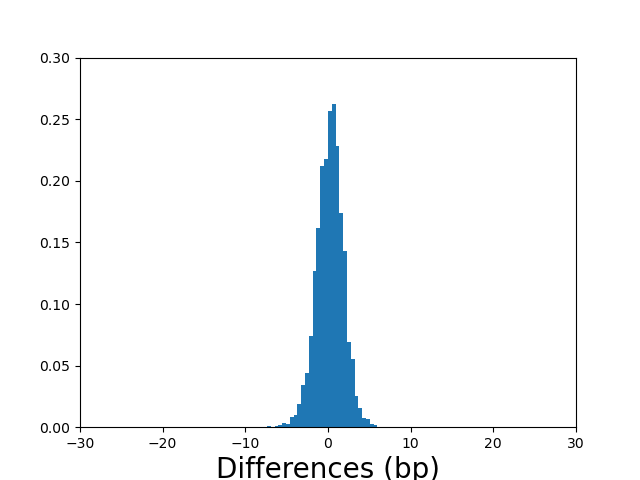}}
			\caption{Pricing differences in basis points. Test Case IV. DANN with joint learning. Columns: $n_f = 2^{22}$ (left), $n_f = 2^{23}$ (central), $n_f = 2^{24}$ (right); Rows: $n_{MC} = 1$ (top), $n_{MC} = 4$ (middle), $n_{MC} = 16$ (bottom).}
			\label{fig:impact_samples_mc_joint}
		\end{figure}

		Next, we perform a more sophisticated experiment with the goal of testing the empirical convergence in two senses: number of samples and Monte Carlo paths per sample. Firstly, we systematically increase the number of samples to measure the impact of this factor in the final estimations. The number of samples used to train the Backward DANNs is now set to $n_b = 2^{23}$, while the number of the samples used to feed the Forward DANN is chosen to be half ($n_f = 2^{22}$), equal ($n_f = 2^{23}$) or double ($n_f = 2^{24}$). Secondly, the number of Monte Carlo paths employed to obtain a sampled payoff is multiplied twice by a factor of 4, i.e, $n_{MC} = 4$ and $n_{MC} = 16$ are tested (besides the single-path original setup). This represents a generalisation of the current approach, making the labels a bit less ``noisy'' and serves to check whether it is worth to arbitrarily increase the training set size or it is better to improve the quality of the labels. Both approaches are combined with the joint learning component. The obtained results are shown in Figures \ref{fig:impact_samples_mc} and \ref{fig:impact_samples_mc_joint} when employing the plain DANN estimator or the DANN plus joint learning, respectively.

		Some important lessons can be extracted from the results in Figures \ref{fig:impact_samples_mc} and \ref{fig:impact_samples_mc_joint}:
		\begin{itemize}
			\item As expected, systematically increase the number of samples provided to the DANN improves the predictions, although the reduction of the interquantile intervals, i.e., in the deviations' variance, is rather limited.
			
			\item In contrast to the previous point, we again observe that the DANN trained relying on the joint learning approach provides more accurate estimations, significantly reducing the variance.
			
			\item The effect of including more Monte Carlo paths per sample presents the expected behavior, i.e., when the number of paths is multiplied by four the error is approximately halved (according to the theoretical convergence rate of Monte Carlo methods, $n^{-1/2}$).

			\item When most of the differences fall below $\pm 3$ basis point, a certain level of saturation is observed meaning that considering either more samples or more Monte Carlo paths per sample no longer reduces the deviations in the predictions (or the reduction results to be negligible).
		\end{itemize}

	\subsection{Valuation at future dates}\label{sec:time}

\begin{figure}[h!]
			\centering
			\subfigure[$t = 8$]{\includegraphics[width=0.49\textwidth]{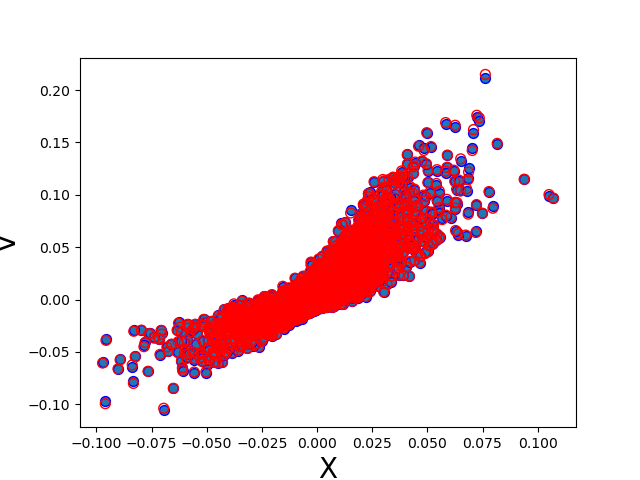}}
			\subfigure[$t = 6$]{\includegraphics[width=0.49\textwidth]{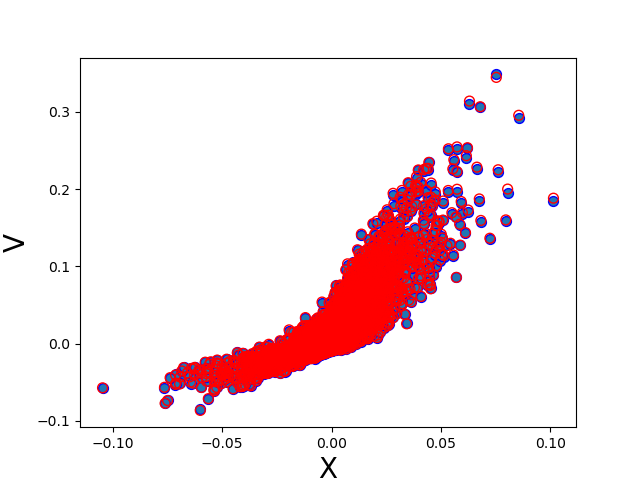}}
			\subfigure[$t = 4$]{\includegraphics[width=0.49\textwidth]{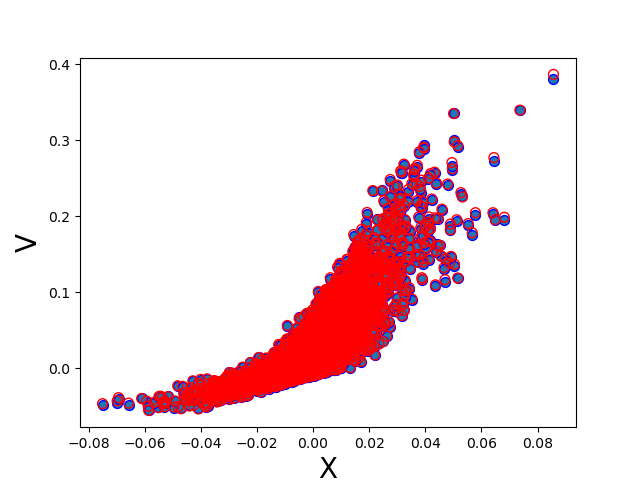}}
			\subfigure[$t = 2$]{\includegraphics[width=0.49\textwidth]{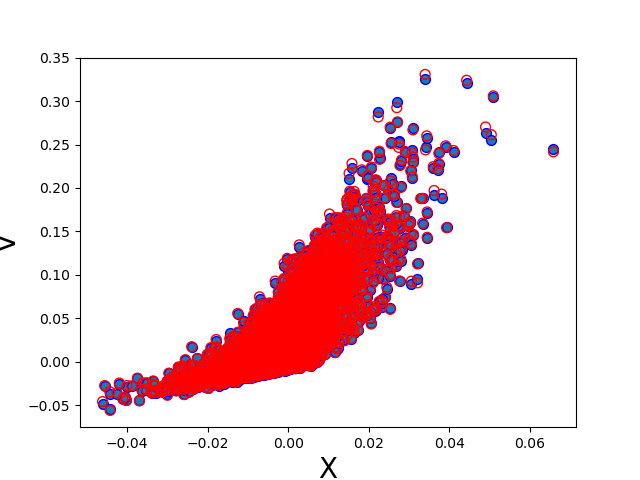}}
			\caption{Predicted prices at different times: Backward DANN (blue points) vs. Forward DANN with time (red circles).}
			\label{fig:CV_Vpred}
		\end{figure}

		The above described methodology can be generalized with the aim of pricing Cancellable IRS not only at the value date (i.e. today's time $t = 0$) but also at some prescribed future dates. For that purpose, the input layer of the forward DANN needs to be enlarged to admit two new inputs: the valuation time $t > 0$ and the state value of LGM process at that particular time instant $x_t$ (which is no longer zero). Furthermore, since two additional inputs are included, a wider DANN is required to manage the increasing complexity of the input's space. Thus, we double the neurons per layer, up to $2^6 = 64$. For simplicity, and without any loss of generality, we only consider the dates coinciding with the cancellation times, assuming that the owner of the product has not cancelled up to the given valuation date. This choice of dates allows us to compare the estimations of the DANN with time component against the trained Backward DANNs. Recall that these network models are trained to estimate the continuation value just after the cancellation times, which is, by definition, the value of the product at that particular time. In Figure \ref{fig:CV_Vpred}, we present the predicted prices by the Backward DANNs and the Forward DANN considering the same time instants and for a range of values of the LGM state variable. We observe that the results are qualitatively very satisfactory, with a high degree of agreement. This also entails a sort of validation for the Backward DANNs.

\begin{figure}[h!]
			\centering
			\subfigure[Abs. avg.: $4.9$, $(Q_{10}, Q_{90}) = (-6.8, 6.9)$.]{\includegraphics[width=0.49\textwidth]{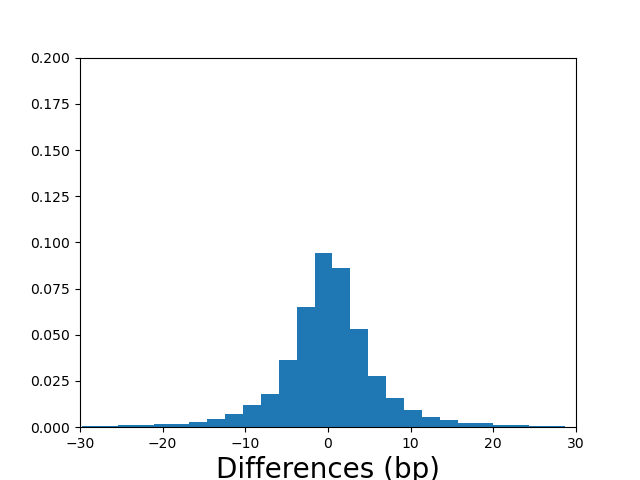}}
			\subfigure[Abs. avg.: $3.8$, $(Q_{10}, Q_{90}) = (-5.7, 3.3)$.]{\includegraphics[width=0.49\textwidth]{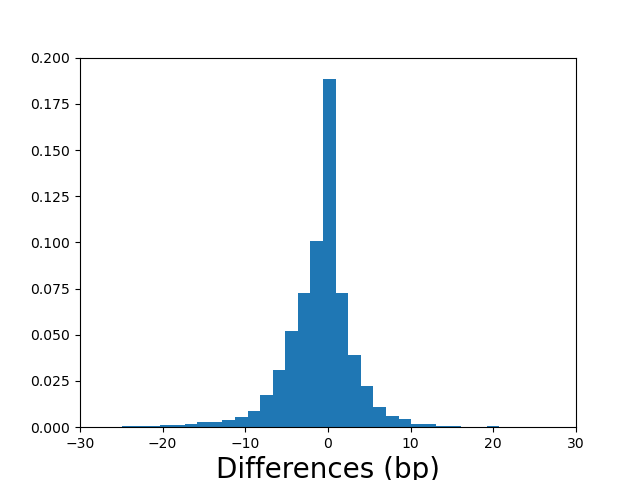}}
			\caption{Pricing differences in basis points, compared against a Monte Carlo pricer: plain DANN (left) and DANN with joint learning (right).}
			\label{fig:diferences_time}
		\end{figure}

		Next, a more quantitative experiment is carried out. We now compare the Forward DANN predictions including the time component with respect to prices computed by means of our Monte Carlo-based pricer (adapted to consider a valuation time different from the value date) which, as mentioned, ensures an error below a half basis point. In this experiment, we also incorporate the joint learning feature, which showed a successful performance in the case without time component. Again, the coterminal European swaptions are the considered aside products, taking into account the valuation time and the LGM state value at that specific time instant i.e., $t$ and $x_t$, respectively. Again, the valuation of these products can be performed analytically, employing the expression \eqref{eq:IRS_European_t}. Note that the prices of the coterminal European swaptions whose maturity is shorter than the handled time are set to zero. In Figure \ref{fig:diferences_time}, the difference distributions and the confidence intervals are shown. Once again, we observe that the DANN including joint learning leads to improved estimations, achieving a reduction in the confidence intervals of around $3$ basis points. In average, the gain in accuracy reaches 1 basis point.

\begin{figure}[h!]
			\centering
			\subfigure[Abs. avg.: $3.1$, $(Q_{10}, Q_{90}) = (-4.2, 4.8)$.]{\includegraphics[width=0.49\textwidth]{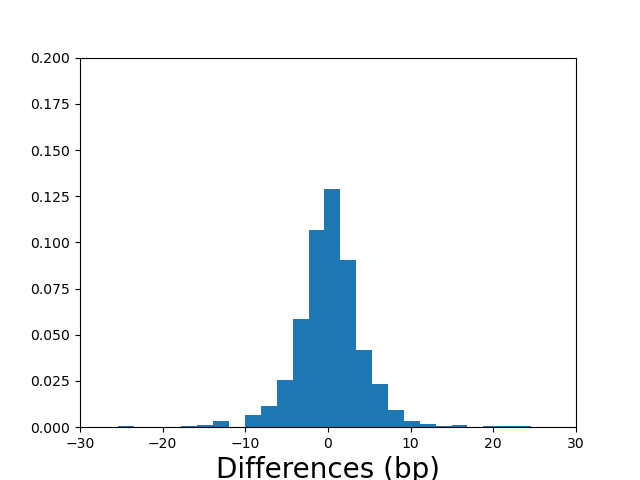}} 
			\subfigure[Abs. avg.: $3.6$, $(Q_{10}, Q_{90}) = (-5.0, 5.9)$.]{\includegraphics[width=0.49\textwidth]{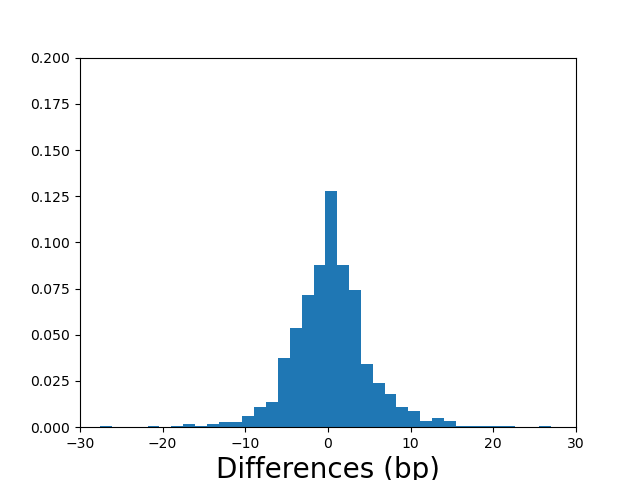}}
			\subfigure[Abs. avg.: $4.5$, $(Q_{10}, Q_{90}) = (-5.4, 6.8)$.]{\includegraphics[width=0.49\textwidth]{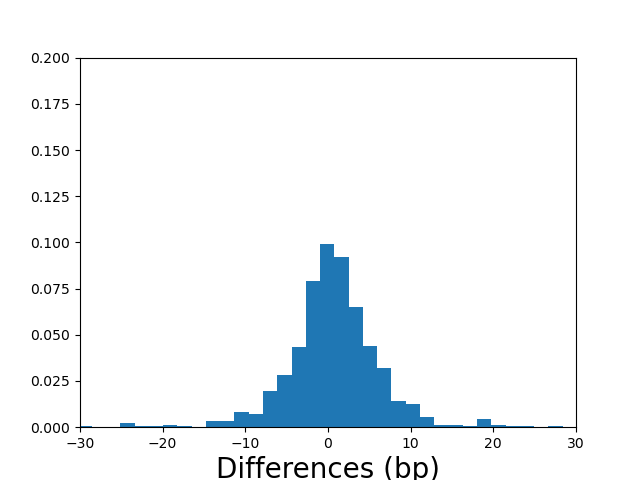}}
			\subfigure[Abs. avg.: $5.9$, $(Q_{10}, Q_{90}) = (-8.5, 7.5)$.]{\includegraphics[width=0.49\textwidth]{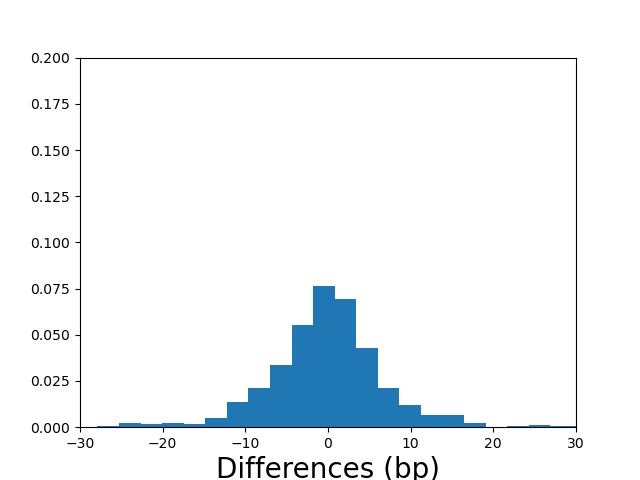}}
			\subfigure[Abs. avg.: $7.9$, $(Q_{10}, Q_{90}) = (-10.1, 10.8)$.]{\includegraphics[width=0.49\textwidth]{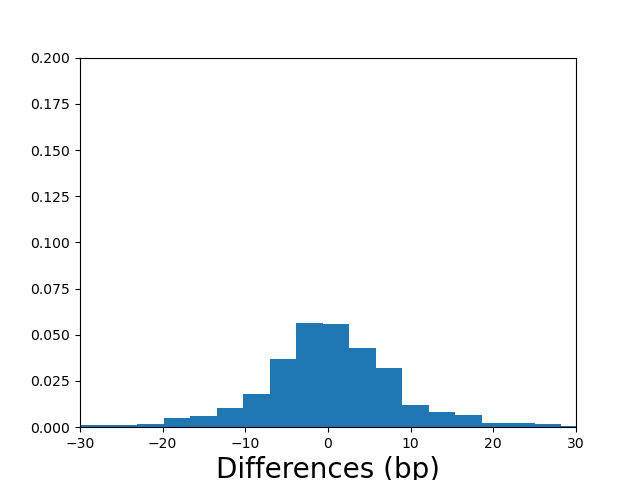}}
			\caption{Pricing differences in basis points, compared against a Monte Carlo pricer, of plain DANN at $t=8$, $t=6$, $t=4$, $t=2$, and $t=0$, from right to left and upper to bottom, respectively.}
			\label{fig:diferences_time_t}
		\end{figure}

\begin{figure}[h!]
			\centering
			\subfigure[Abs. avg.: $2.2$, $(Q_{10}, Q_{90}) = (-3.3, 3.2)$.]{\includegraphics[width=0.49\textwidth]{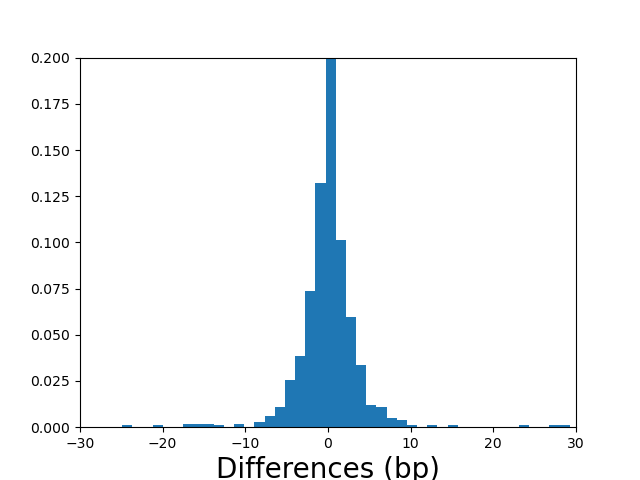}} 
			\subfigure[Abs. avg.: $2.8$, $(Q_{10}, Q_{90}) = (-3.7, 3.9)$.]{\includegraphics[width=0.49\textwidth]{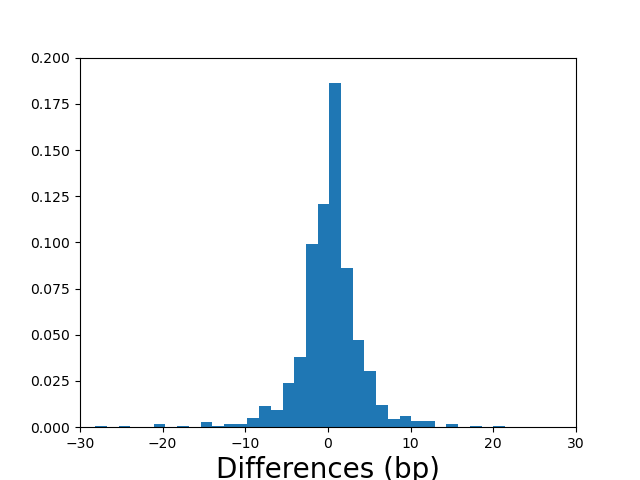}}
			\subfigure[Abs. avg.: $3.1$, $(Q_{10}, Q_{90}) = (-5.3, 2.9)$.]{\includegraphics[width=0.49\textwidth]{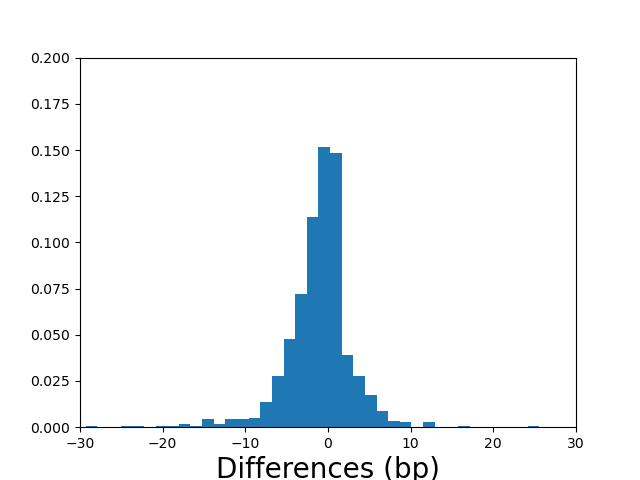}}
			\subfigure[Abs. avg.: $3.7$, $(Q_{10}, Q_{90}) = (-6.9, 2.8)$.]{\includegraphics[width=0.49\textwidth]{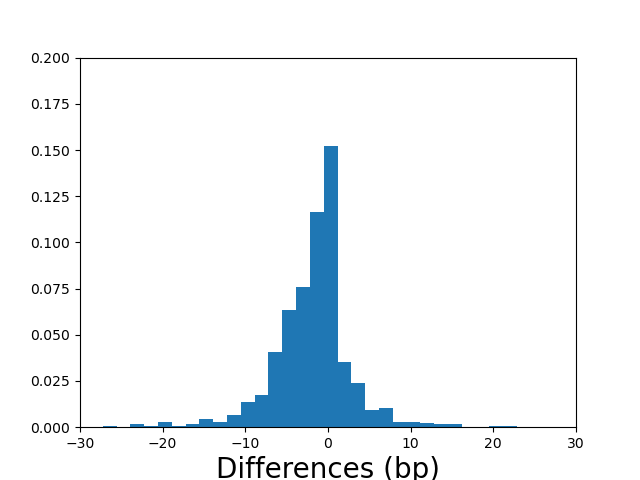}}
			\subfigure[Abs. avg.: $3.7$, $(Q_{10}, Q_{90}) = (-7.3, 3.4)$.]{\includegraphics[width=0.49\textwidth]{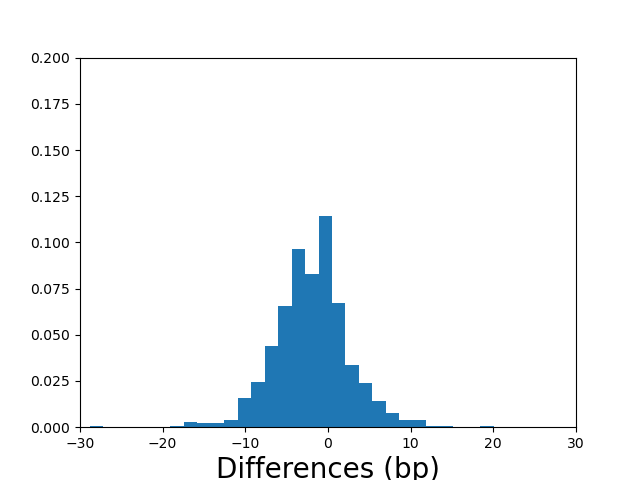}}
			\caption{Pricing differences in basis points, compared against a Monte Carlo pricer, of  DANN with joint learning at $t=8$, $t=6$, $t=4$, $t=2$, and $t=0$, from right to left and upper to bottom, respectively.}
			\label{fig:diferences_time_joint_t}
		\end{figure}

		The previous results are disaggregated in terms of the considered valuation times. The obtained differences' distributions are presented in Figures \ref{fig:diferences_time_t} and \ref{fig:diferences_time_joint_t}.
		This disaggregated view offers more relevant information, where the impact of the joint learning in the estimation is much clearer. In the following, we highlight some of the observed outcomes:
		\begin{itemize}
			\item More accurate estimations for times closer to the end of the derivative contract are obtained, which is explained by the much lower variability/complexity of the predicted prices when the number of remaining payments in the underlying IRS is small. As we move to times closer to zero, the pricing problem becomes more challenging and, consequently, the DANN predictions deteriorate.

			\item The inclusion of the joint learning approach in the DANN fitting process provides again a remarkable improvement in the estimations, having more positive impact for shorter times where, as mentioned above, more error has been observed when the plain DANN is employed. This very desirable effect represents another relevant advantage provided by the joint learning strategy.

			\item The histograms of differences in the estimations provided by the DANN with joint learning present fatter left tails, thus indicating that this ANN model tends to underestimate the correct value. This effect appears due to the fact that the coterminal European swaptions are, by definition, cheaper than the product at hand (cancelable IRS), pushing the estimations down. This is more pronounced for shorter times, where the contribution of the coterminal swaptions is more important (the higher the time, the more aside products with zero value), resulting in histograms that present a deeper ``unbiased'' pattern.

		\end{itemize}

\section{Conclusions}\label{sec:conclusions}

    In this work, an innovative solution for the Bermudan swaption valuation problem based on advanced deep learning techniques has been proposed. In this setting, many additional very relevant components have been added on top of classical ANN approach. Some of them are appropriate adaptations of existing methodologies, like the use of sampled payoffs (highly noisy price values) as labels or the differential machine learning network design. Those features have been carefully tested and analyzed in a challenging and practical financial problem, as it is the case of pricing Bermudan swaptions. Moreover, we have also proposed a novel training strategy in quantitative finance, namely the joint learning, which has shown an impressive performance. The idea behind our joint learning approach is to incorporate ``similar'' financial products as outputs, aiming that they help in the training process to reach more accurate solutions for complex derivatives at less/similar computational cost. More precisely, for pricing Bermudan swaptions, we consider the equivalent pricing problem of pricing cancellable IRS and we choose a set of European swaptions with maturities at the cancellation times as additional outputs for the joint learning strategy. Throughout several experiments, the advantages of employing the proposed joint learning-based training have been clearly highlighted.

    Among the conclusions of this work, we also note that application of the set of proposed techniques can be extended to the pricing of other financial products, such as autocallables on a basket of assets, for which the Greeks can also be obtained. Also in this case, the joint learning strategy can be applied by selecting the characteristics of the appropriate additional financial products. Some research work in this direction has been already initiated by the authors.

\section*{Acknowledgements}

    The authors AL and CV acknowledge the support of Centre for Information and Communications Technology Research (CITIC). CITIC is funded by the Xunta de Galicia through the collaboration agreement between the Consellería de Cultura, Educación, Formación Profesional e Universidades and the Galician universities for the reinforcement of the research centres of the Galician University System (CIGUS). This research has been mainly funded under a contract between BBVA and CITIC. The contents of this article represent only the authors’ views, and do not represent the opinions of any firm or institution.

\bibliographystyle{plain}
\bibliography{swaption.bib}

\end{document}